\documentclass[11pt,letterpaper]{article}

\usepackage{jheppub}
\usepackage{color}
\usepackage{graphicx}
\usepackage{wrapfig,enumerate,slashed}
\usepackage{caption}
\usepackage{subcaption}
\usepackage{multirow}

\usepackage{footmisc}
\usepackage{amsmath}
\usepackage{wasysym} 
\usepackage{comment}
\usepackage{hyperref}
\usepackage{makecell}

\newcommand{\rig}{\rightarrow}

\newcommand{\be}{\begin{eqnarray}}
\newcommand{\ee}{\end{eqnarray}}

\newcommand{\bee}{\begin{eqnarray}}
\newcommand{\eee}{\end{eqnarray}}
\newcommand{\beeq}{\begin{equation}}
\newcommand{\eeeq}{\end{equation}}

\newcommand{\mc}{\mathcal}
\newcommand{\mr}{\mathrm}

\newcommand{\LT}{\text{T}}
\definecolor{red}{rgb}{1,0,0}

\definecolor{purple}{rgb}{0.5,0,0.5}

\newcommand{\GeV}{\text{~GeV}}
\newcommand{\TeV}{\text{~TeV}}

\begin{document}

\title{Quark-Gluon tagging with Shower Deconstruction: Unearthing dark matter and Higgs couplings}

\author[a]{Danilo Ferreira de Lima,}
\author[b]{Petar Petrov,}
\author[c]{Davison Soper}
\author[b]{and Michael Spannowsky}

\affiliation[a]{Physikalisches Institut, Ruprechts-Karls-Universit\"at Heidelberg, 69120 Heidelberg, Germany}
\affiliation[b]{Institute for Particle Physics Phenomenology, Department of Physics, Durham University, DH1 3LE, United Kingdom}
\affiliation[c]{Institute of Theoretical Science University of Oregon
Eugene, OR 97403-5203, USA}

\emailAdd{dferreir@cern.ch}
\emailAdd{p.m.petrov@durham.ac.uk}
\emailAdd{soper@uoregon.edu}
\emailAdd{michael.spannowsky@durham.ac.uk}

\abstract{The separation of quark and gluon initiated jets can be an important way to improve the sensitivity in searches for new physics or in measurements of Higgs boson properties. We present a simplified version of the shower deconstruction approach as a novel observable for quark-gluon tagging. Assuming topocluster-like objects as input, we compare our observable with energy correlation functions and find a favorable performance for a large variety of jet definitions.  We address the issue of infrared sensitivity of quark-gluon discrimination. When this approach is applied to dark matter searches in mono-jet final states, limitations from small signal-to-background ratios can be overcome. We also show that quark-gluon tagging is an alternative way of separating weak boson from gluon-fusion production in the process $p + p \to H + \mathrm{jet} + \mathrm{jet} + X$. }

\preprint{DCPT/16/140, IPPP/16/70}

\maketitle


\section{Introduction}
\label{sec:intro}

Quark-gluon tagging of jets can be an important tool to separate signal from backgrounds. For instance, it is of interest to search for dark matter production by using the process in which produced dark matter particles recoil against a single jet, as described in \cite{Khachatryan:2014rra}. Particularly when the mediator between the dark matter and the Standard Model particles is a scalar that couples preferably to the third-generation fermions, the associated jet is likely to be gluon initiated. One of the dominant Standard Model backgrounds however is the production of a jet plus a Z boson, in which the Z boson decays to $\nu \bar \nu$. In the tree level diagram for the background, the jet can be either quark initiated or a gluon initated. Thus if we can preferentially reject quark jets and keep gluon jets, we can improve the ratio of signal events retained to background events retained.

Conversely, many measurements of Higgs boson properties and couplings rely on the weak-boson-fusion production process $qq \to Hqq$ \cite{Dokshitzer:1991he,Rainwater:1998kj,Zeppenfeld:2000td,Englert:2015hrx}. In particular, if one wants to measure the Higgs boson coupling to gauge bosons, one wants to look at this process and not the dominating gluon-fusion process $gg \to Hgg$ \cite{DelDuca:2001fn}. In $qq \to Hqq$, there are two quark jets, while in $gg \to Hgg$, there are two gluon jets. Hence, here we would prefer to reject gluon jets and keep quark jets to improve the precision of the measurement.

A third example would be the decays of squarks into jets and the lightest supersymmetric particle. Heavy squarks of the first and second generation decay almost exclusively into quarks and gauginos, while jets and missing transverse energy (MET) backgrounds have a larger gluon-jet component.

In all examples above, exploiting the different admixture of gluon and quark initiated jets can help to improve the signal-to-background ratio. Consequently, several observables have been proposed to exploit the differences in the radiation profiles of quarks and gluons \cite{Catani:1992jc,Thaler:2010tr,Gallicchio:2011xq,Larkoski:2013eya,Larkoski:2014pca,Bhattacherjee:2015psa,Badger:2016bpw} and have been studied in data by ATLAS \cite{Aad:2014gea} and CMS \cite{CMS-PAS-JME-13-002}.

Suppose that we want to accept quark jets and reject gluon jets. Typically, one can adjust the parameters of the algorithm we use so as to obtain a desired fraction $\varepsilon_s$ of quark jets accepted. Then $\varepsilon_b^{-1}$, the inverse of the fraction of gluon jets accepted, will depend on $\varepsilon_s$. In this paper, we present ``ROC'' curves showing $\varepsilon_b^{-1}(\varepsilon_s)$ versus $\varepsilon_s$. We want $\varepsilon_b^{-1}$ to be as large as possible for any given $\varepsilon_s$. However, this performance metric is not the only issue that we need to address. We also need to know with reasonable accuracy the value of $\varepsilon_b^{-1}(\varepsilon_s)$ for a given $\varepsilon_s$. This information can come from experiment if the function $\varepsilon_b^{-1}(\varepsilon_s)$ is characteristic of quark-initiated versus gluon-initiated jets independently of how the jets are produced. We will investigate whether this is so in section \ref{sec:systematics}. Information on $\varepsilon_b^{-1}(\varepsilon_s)$ for a given tagging method can also come from perturbation theory and simulation using parton shower event generators. Here, the findings of \cite{Aad:2014gea} indicate the need for the inclusion of certain detector effects in phenomenological analyses and the benefit of observables that are largely insensitive to non-perturbative effects. In this paper, we try to avoid sensitivity to parton splitting processes at very small momentum scales. For instance, we use observables that are technically infrared safe. However, we will discover that it is precisely parton splitting processes at quite small momentum scales that best distinguish the substructure of a quark jet from that of a gluon jet. Thus we cannot avoid a certain degree of infrared sensitivity. We return to this issue in section \ref{sec:systematics}.

In this paper, we explore the use of several methods to distinguish between quark in gluon jets in $p + p \to Z + \mathrm{jet} + X$ and $p + p \to \mathrm{jet} + \mathrm{jet} + X$ events. We evaluate the performance and simulation uncertainties of the shower deconstruction method \cite{Soper:2011cr,Soper:2012pb,Soper:2014rya} and compare it to the use of energy correlation functions \cite{Larkoski:2013eya}. 

The structure of the paper is as follows: In section~\ref{sec:substructure} we describe our analysis setup and the algorithms applied for quark/gluon tagging, emphasizing a method based on shower deconstruction. In section~\ref{sec:tag}, we discuss their performance and uncertainties of these algorithms. We apply quark/gluon tagging based on shower deconstruction to dark matter searches and $p + p \to  H + \mathrm{jet} + \mathrm{jet} + X$ production and evaluate by how much the signal-to-background ratio can be improved in section~\ref{sec:application}. In section~\ref{sec:conclusions} we offer a summary and our conclusions.

\section{Jet substructure for quark-gluon tagging}
\label{sec:substructure}

In this section, we first describe the analysis setup for the paper. Then we discuss the input objects that we use for quark-jet versus gluon-jet discrimination.  Next, we turn to the observables that we use.

\subsection{The analysis setup}
\label{sec:setup}

Our aim in this paper is to test the performance of algorithms designed to discriminate between quark-initiated jets and gluon-initiated jets. For this, we use two types of of events generated using Pythia 8 \cite{Sjostrand:2007gs} with initial state radiation and underlying event switched on. The first type, and the one on which we will focus most, is a single jet with an associated invisible Z boson - $qg\rig qZ(\nu\bar{\nu})$, $q\bar{q}\rig gZ(\nu\bar{\nu})$. The other, which we use to show how much the tagging efficiency is affected by the event color flow, is dijet production $qq/gg\rig qq$, $q\bar{q}/gg\rig gg$. We generate four sets of each type in order to compare the performance at different limits for the transverse momentum in the hard scattering: $p_T > 200, 400, 600, 1000$ GeV. 

For each event, we begin with input objects. The input objects can be hadrons, tracks, or certain calorimeter based objects, as described in the following subsection. We cluster the input objects into jets and select the leading jet: the one with the greatest transverse momentum. This is the ``fat jet'' that we wish to tag as being a probable quark jet or a probable gluon jet. To proceed, there should be at least one jet in the rapidity range $|y|<5$ for $Z+\mathrm{jet}$ events or two such jets for $\mathrm{dijet}$ events. For the clustering into jets, we use the C/A algorithm with a standard radius $R_\mr{fj}=0.4$ and a transverse momentum that reflects the event generation limit $p_{T\mr{fj}}>p_{T\mr{limit}}$. With $R = 0.4$, the fat jet is not so fat. This choice follows from the fact that we are analyzing the QCD radiation in the jet rather than looking for the decay of a heavy particle as is the case in many jet substructure studies. We also use a larger radius jet definition at $R_\mr{fj}=0.8$ for some analyses. 

\subsection{Input objects}
\label{sec:inputs}

The observable quantities that we analyze for their ability to distinguish quark jets from gluon jets are built from certain input objects. We study four different classes of input objects: hadrons, tracks, and two sorts of calorimeter based objects. 

While hadrons as input objects provide the most detailed information in the substructure of a jet, they are unlikely to be accessible in an experimental environment. 

Using tracks allows very good angular resolution, but only for charged particles, while being blind to neutral particles. For tracks, we do not include a detector simulation, so that we do not take into account track efficiencies or energy smearing of tracks. Thus we likely overestimate the performance of the observables with track inputs. 

Most of the analyses that we present are based on input objects built from idealized calorimeter cells. In general purpose experiments such as ATLAS~\cite{atlas} and CMS~\cite{cms}, often the calorimeter cells are not directly used to make jets. Instead, a combination of cells is used. 

ATLAS uses ``topoclusters"~\cite{Aad:2011he,Lampl:2008zz, topoclusters}. 
A topocluster is a group of topologically connected calorimeter cells, which
are chosen based on an algorithm to suppress calorimeter noise. The
algorithm starts by choosing a ``seed" calorimeter cell, which has a signal over
noise ratio over a specific threshold. It then combines it with neighbour cells
that satisfy a minimum signal-to-noise ratio criterion iteratively.
This method improves the jet algorithm inputs signal-to-noise ratio.
Although it has the positive effect of improving the calorimeter's
signal-to-noise ratio~\cite{topoclusters}, it imposes a limitation in the angular resolution of the experiments. While the algorithm used to create topoclusters is clearly defined, the angular resolution limitation is not explicit in the algorithm. It depends on the calorimeter's noise average and cell sizes, which vary in both ATLAS and CMS, depending on the jet position.

Following a somewhat different approach, CMS uses so-called particle-flow (PF) objects \cite{CMS-PAS-PFT-09-001}. PF objects consist of all visible particles in an event, i.e. muons, electrons, photons, charged hadrons, and neutral hadrons. Charged hadrons, electrons and muons are predominantly reconstructed from tracks in the tracker, while photons and neutral hadrons are reconstructed from energy deposits in topoclusters. Combining the topocluster and tracking system information, CMS can greatly improve the PF jets' spatial resolution with respect to calorimeter jets, e.g. by exploiting tracking information \cite{Katz:2010mr,Schaetzel:2013vka,Larkoski:2015yqa,Spannowsky:2015eba}. However, the jet-energy-resolution deteriorates quickly for jets with $R \leq 0.2$ \cite{partflow}. Hence, the way CMS uses its PF objects currently results in a lower limit on the spatial resolution of jets, just as the angular resolution is limited by the size of topoclusters in ATLAS.

We conclude that jet substructure methods must take into account the finite angular resolution of calorimeter objects used as substructure inputs. In this phenomenological study, we approximate this resolution limitation by using Cambridge-Aachen (CA) \cite{Dokshitzer:1997in,Wobisch:1998wt} jets with an $R$ parameter of 0.1 and $p_T>1$ GeV as input to the algorithms. We use two sorts of calorimetric input objects, which we call ``massive topoclusters'' and ``massless topoclusters.''

ATLAS topoclusters are forced to be massless. That is, after measuring the energy, pseudorapidity and azimuthal angle of the topocluster, its three-momentum is scaled to create a vector with $p^2 = 0$. We create massless topoclusters with this rescaling. However, we mostly use massive topoclusters, in which the topocluster momentum $p$ is the sum of the momenta of the constituent particles, so that $p^2 > 0$. 

We mostly use massive instead of massless input objects because we find that neglecting their masses leads to a deterioration in quark-gluon discrimination. One could imagine using a similar procedure to that described in \cite{ATLAS-CONF-2016-035} to calibrate the masses of small jets, analogous to our ``massive topoclusters.''

\subsection{Observables for quark-gluon tagging}
\label{sec:observables}

We will use two classes of jet substructure observables in order to distinguish quark jets from gluon jets. One is based on shower deconstruction, the other is based on energy correlations. We begin with shower deconstruction.

\subsubsection{Shower deconstruction}
\label{sec:sd}

Shower deconstruction \cite{Soper:2011cr, Soper:2012pb, Soper:2014rya} is a general method for distinguishing events created by a sought signal process from events created by other, less interesting, processes. In this case, the ``signal'' process creates a quark-initiated jet and we wish to distinguish this quark jet from ``background'' gluon jets. (Of course, we could reverse the roles of signal and background here.) We start with a list of the momenta $\{p\}_m = \{p_1, p_2,\dots,p_m\}$ of $m$ microjets -- small radius jets -- constructed from the contents of the larger fat jet. We calculate an approximation $P(\{p\}_m|q)$ that the observed microjets could be the result of a parton shower that starts with a quark parton and ends with $m$ partons with momenta $\{p\}_m$. We similarly calculate an approximate probability $P(\{p\}_m|g)$ to obtain the observed microjets starting from a quark. Then we form the likelihood ratio
\begin{equation}
\label{eq:chidef}
\chi(q,g) = \frac{P(\{p\}_m|q)}{P(\{p\}_m|g)}
\;,
\end{equation}
where the first argument indicates the signal hypothesis, {\it i.e.}\ quarks, and the second argument the background hypothesis, {\it i.e.}\ gluons. Note that $\chi(g,q) = 1/\chi(q,g)$. A large value of $\chi(q,g)$ indicates a likely quark jet, while a small value of $\chi(q,g)$ indicates a likely gluon jet. Thus imposing a cut $\chi(q,g) > \chi_{\rm cut}$ tags quark jets and imposing a cut $\chi(g,q) > \chi_{\rm cut}$ tags gluon jets.

The idea of the shower deconstruction method here is to distinguish the radiation pattern created by an initial quark from the radiation pattern of a gluon. This is rather different from our previous applications of shower deconstruction, in which the aim is to distinguish the pattern of partons produced by the decay of a heavy particle, such as a top quark, from the pattern of partons produced by normal QCD radiation. Distinguishing quark jets from gluon jets is harder. We have normal QCD radiation in either case, but gluon jets have, on average, more radiation because gluons have a larger color charge. We expect to see two differences between quark and gluon jets. First, gluon jets ought to be more likely to contain more microjets than quark jets. Second, the virtuality $p_i^2$ of the highest $p_\LT$ microjet is likely to be larger in the gluon case than in the quark case because the microjet contains more radiation inside it even though the radiation is clustered into a single microjet.

\begin{figure} 
\centering
\includegraphics[width=1.1\linewidth]{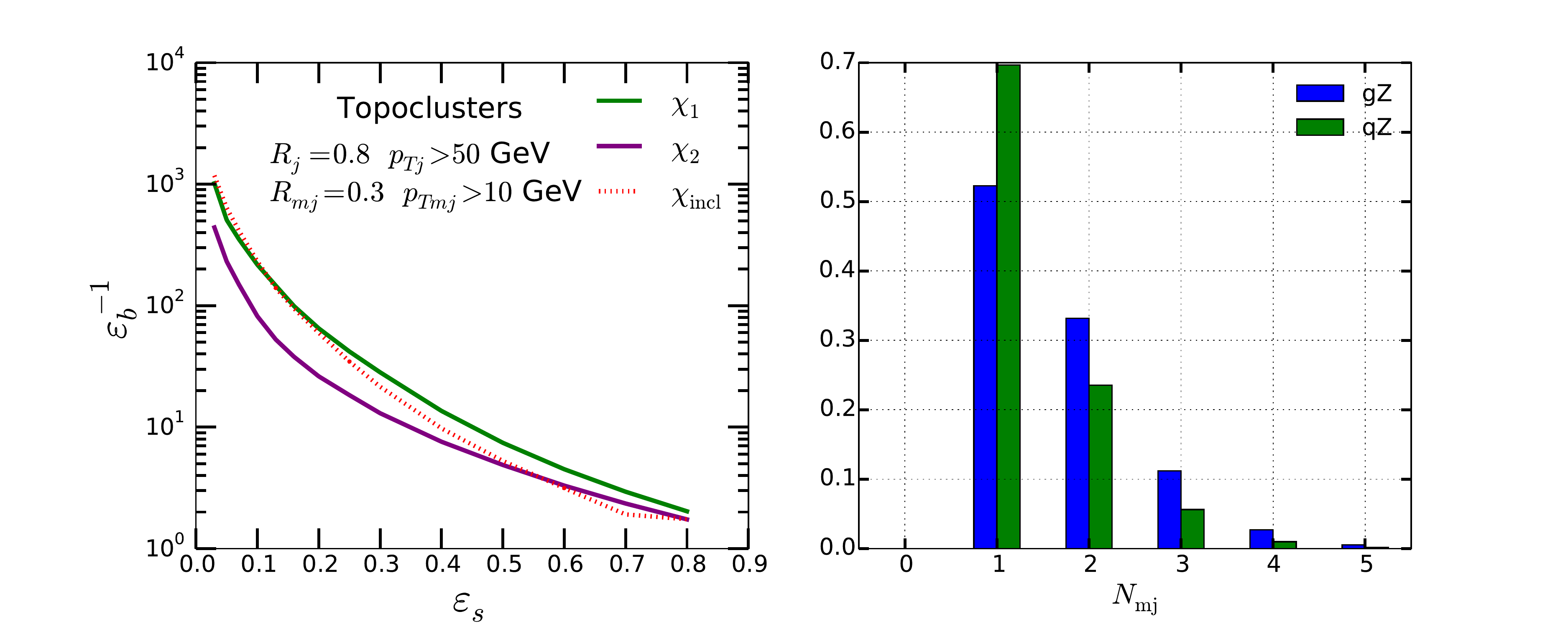}
\caption{ Left: quark (signal) vs gluon (background) ROC curves for $\chi$ with exactly one or exactly two microjets. Right: microjet multiplicity distribution.}
\label{fig:nSD}
\end{figure}

To see how this works, we apply shower deconstruction for $qg\rig qZ(\nu\bar{\nu})$, $q\bar{q}\rig gZ(\nu\bar{\nu})$ events, taking massive topoclusters as the the input objects and using them to define a fat jet using a jet radius $R = 0.8$. The massive topoclulsters in the original fat jet are grouped into microjets using the $k_\LT$ algorithm with radius $R_\mr{mj} = 0.3$ and a minimum transverse momentum $p_{T\mr{mj}}^{\mr{min}} = 10 \GeV$. Then the likelihood variable $\chi$ from eq.~(\ref{eq:chidef}) is calculated for each event. Different events have different numbers of microjets. In the right hand plot of figure~\ref{fig:nSD}, we plot the number of microjets in the $gZ$ sample (blue) and in the $qZ$ sample (green). Not surprisingly, quark jets are more likely than gluon jets to produce just one microjet, while gluon jets produce more microjets. This feature can help distinguish quark jets from gluon jets. However, when we look at the distribution of $\chi$ for those events with exactly one microjet, we find better quark-gluon discrimination than when we look for  $\chi$ for those events with exactly two microjets, as illustrated in the left-hand plot of figure \ref{fig:nSD}. This suggests that there is a lot of discriminating power in the shower-deconstruction $\chi$ for the simple case of one microjet. In fact, we find that when we simply calculate $\chi$ for the fat jet as a whole, without decomposing it into microjets, we get quark-gluon discriminating power that is often better than when the fat jet is decomposed into several microjets. This behavior is in sharp contrast to applications in which one wants to distinguish  ordinary QCD jets from jets arising from the decay of a heavy particle like a top quark: it is important that a top quark decays into at least three jets.

Because using shower deconstruction with just one microjet works quite well, it is of interest to understand what shower deconstruction does in this case. The formula for $\chi$ for just one microjet is simply a ratio of Sudakov factors:
\begin{equation}
\label{eq:SDo}
\chi = \frac{P(\{p\}_m|q)}{P(\{p\}_m|g)} = \frac{e^{-\mc{S}_q}}{e^{-\mc{S}_g}} = 
e^{-\left(\mc{S}_\mr{qqg}\Theta\left(\mc{S}_\mr{qqg}>0\right) - \mc{S}_\mr{ggg}\Theta\left(\mc{S}_\mr{ggg}>0\right) - n_f\mc{S}_\mr{gqq}\right)}
\;.
\end{equation}
where 
\begin{equation}
\label{eq:SD1}
\begin{split}
&\mc{S}_\mr{qqg} = \frac{C_\mr{F}}{\pi b_0^2}\left\{ \mr{ln}\left(\frac{\alpha_\mr{S}(\mu_J^2)}{\alpha_\mr{S}(k_J^2)}\right)\left[\frac{1}{\alpha_\mr{S}(R_{\rm fj}^2k_J^2)} - \frac{3b_0}{4}\right] + \frac{1}{\alpha_\mr{S}(\mu_J^2)} - \frac{1}{\alpha_\mr{S}(k_J^2)} \right\}\;,\\\\
&\mc{S}_\mr{ggg} = \frac{C_\mr{A}}{\pi b_0^2}\left\{ \mr{ln}\left(\frac{\alpha_\mr{S}(\mu_J^2)}{\alpha_\mr{S}(k_J^2)}\right)\left[\frac{1}{\alpha_\mr{S}(R_{\rm fj}^2k_J^2)} - \frac{11b_0}{12}\right] + \frac{1}{\alpha_\mr{S}(\mu_J^2)} - \frac{1}{\alpha_\mr{S}(k_J^2)} \right\}\;,\\\\
&\mc{S}_\mr{gqq} = \frac{T_\mr{R}}{3\pi b_0}\mr{ln}\left(\frac{\alpha_\mr{S}(\mu_J^2)}{\alpha_\mr{S}(k_J^2)}\right)
\;.
\end{split}
\end{equation}
Here $\mu_J$ is the jet mass, $k_J$ is the jet transverse momentum, and $b_0 = (33 - 2 n_{\rm f})/(12 \pi)$.

In the case that we evaluate $\chi$ with simply the whole jet as the single microjet, we see that $\chi$ is a function of only two variables, the jet mass $\mu_J$ and the jet transverse momentum $k_J$. The function $\ln\chi$ is an approximation to the likelihood ratio
\begin{equation*}
\ln{L(q,g)} = \ln{P_\mr{MC}(\mu_J^2,k_J^2|q)} - \ln{P_\mr{MC}(\mu_J^2,k_J^2|g)}
\;.
\end{equation*}
If we use only the two variables $\mu_J^2$ and $k_J^2$ to describe fat jets in each event, then $\ln{L(q,g)}$ provides the optimum way to distinguish quark jets from gluon jets as long as $P_\mr{MC}(\mu_J^2,k_J^2|q)$ and $P_\mr{MC}(\mu_J^2,k_J^2|g)$ provide accurate representations of nature. Thus one way to test whether the shower deconstruction variable $\chi$ is doing a good job is to construct the $\ln{L(q,g)}$ and compare $\ln \chi$ to $\ln{L(q,g)}$.

\begin{figure} 
\centering
\includegraphics[width=0.49\linewidth]{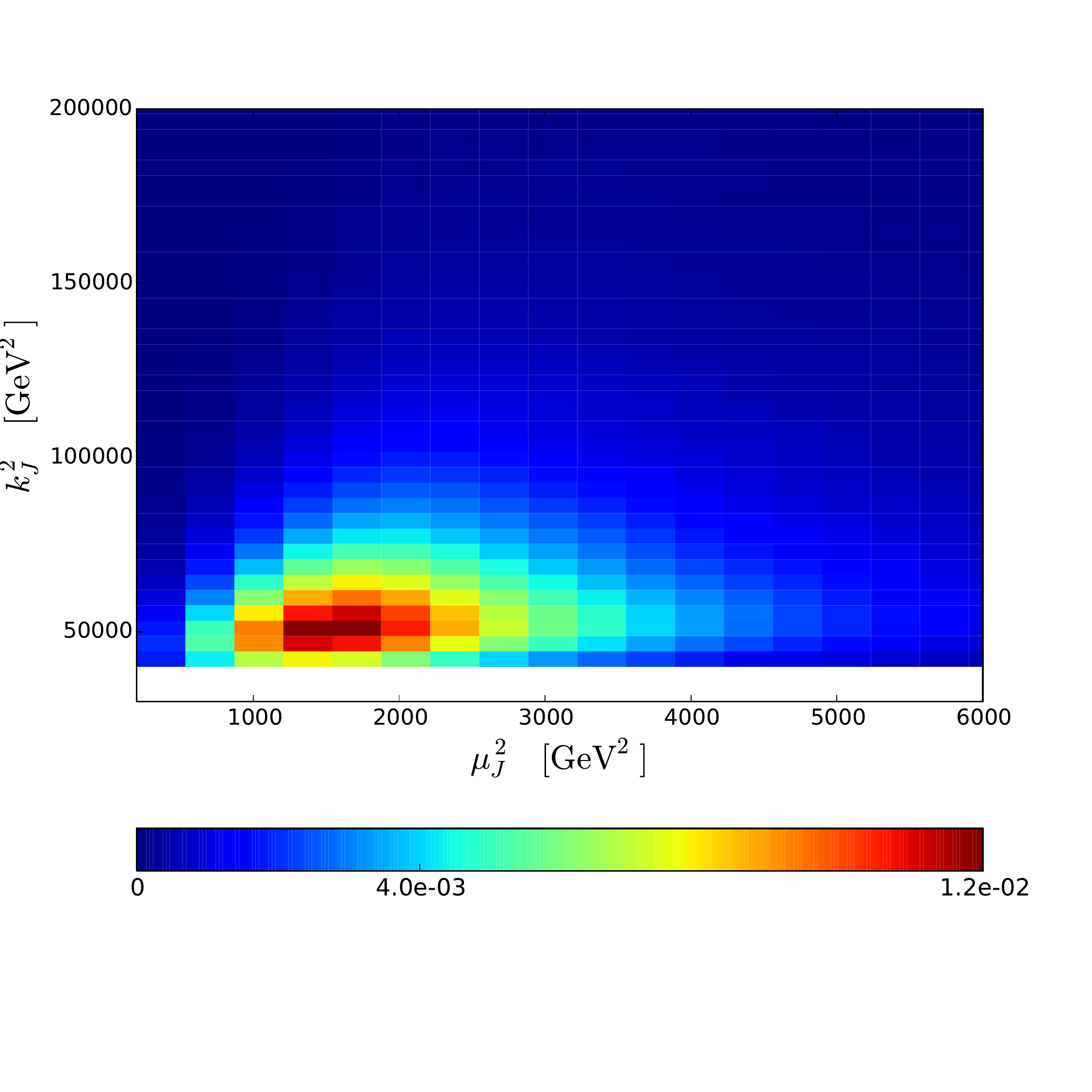}
\includegraphics[width=0.49\linewidth]{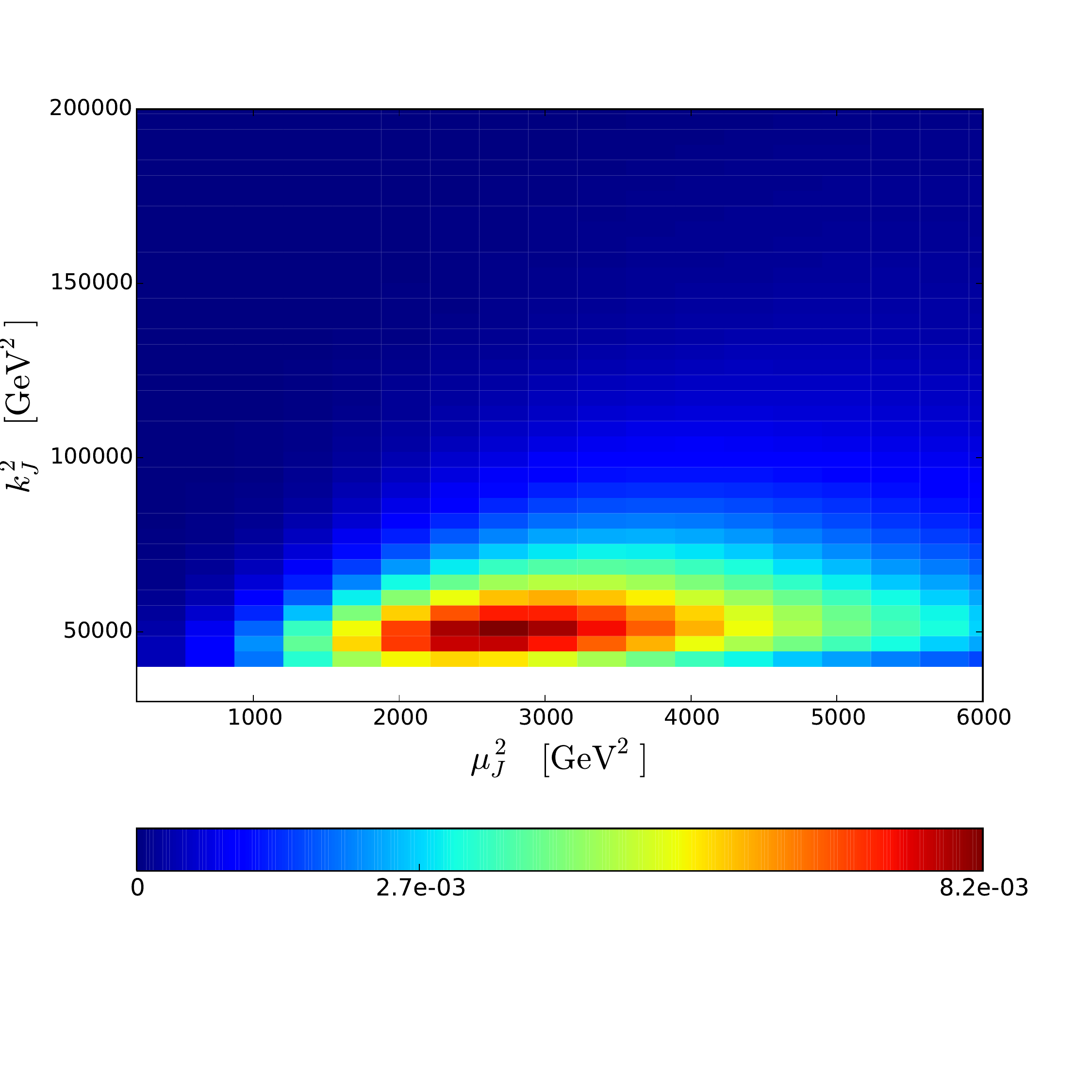} \\
\includegraphics[width=0.49\linewidth]{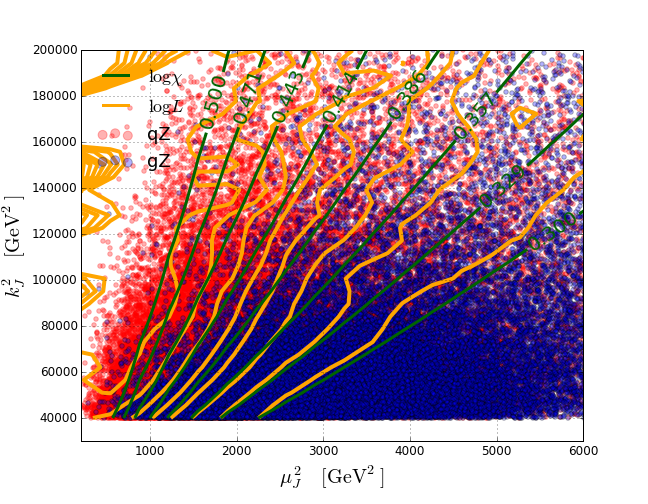}
\caption{ Gaussian kernel-density estimate of the $R=0.4$ leading jets' mass and transverse momentum distribution in $Z+q$ (left) and $Z+g$ (right) events. In the bottom plot we overlay a scatter plot of the two distributions, contours of the likelihood derived from the gaussian kernel-density estimator and another contour plot of the shower deconstruction variable $\chi$.}
\label{fig:Contour_kde}
\end{figure}

To build the likelihood function $L(q,g)$, we use the normalized $(\mu_J^2,k_J^2)$ histogram for the leading jets in $Z+q$ and $Z+g$ events. Then the likelihood in each bin is the ratio of the probability between the quark and gluon samples for that bin. However, the latter are strongly influenced by statistical fluctuations. We attempt to ameliorate this by ``spreading" the probability of each bin. We use the gaussian kernel-density estimator \cite{ScottKDE} to smear the probability contained in each bin into a 2-dimensional gaussian distribution with the same normalization. The volume and mean of the gaussian kernel is fixed by the data, but the standard deviation is a free  parameter that determines the ``smoothing" effect. Even though the best way to determine this bandwidth parameter is through a cross-validation metric, we choose the parameter by visual comparison with the histograms. This leads to the distributions and contours in figure \ref{fig:Contour_kde}. The axes represent our  two variables, $\mu_J^2$ and $k_J^2$. In the bottom figure, we overlay three plots. The first is a scatter plot for the events in $Z+q$ jets and in $Z+g$ jets. The second, in yellow, is plot  of contour lines of $\ln{L(q,g)}$ (after smoothing as described above). The third, in green, is a plot of contour lines of $\ln \chi$. We conclude that $\ln \chi$ is a reasonably good approximation to $\ln{L(q,g)}$.

In the analyses that follow, we mostly apply shower deconstruction to smaller, $R = 0.4$, fat jets, taking massive topoclusters as the input and using just one microjet, which is then equal to the whole fat jet.

\begin{figure} 
\centering
\includegraphics[width=0.49\linewidth]{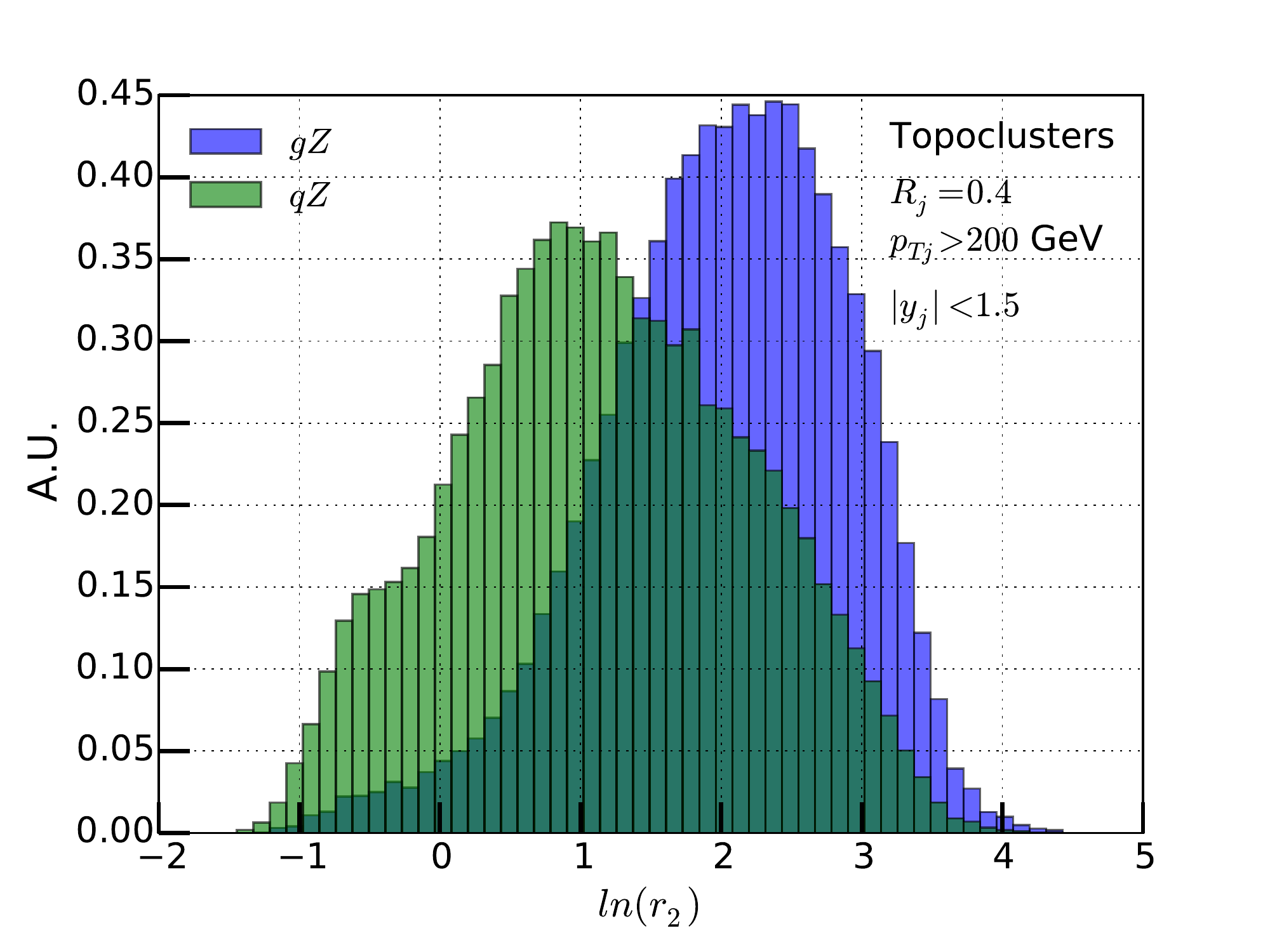}
\includegraphics[width=0.49\linewidth]{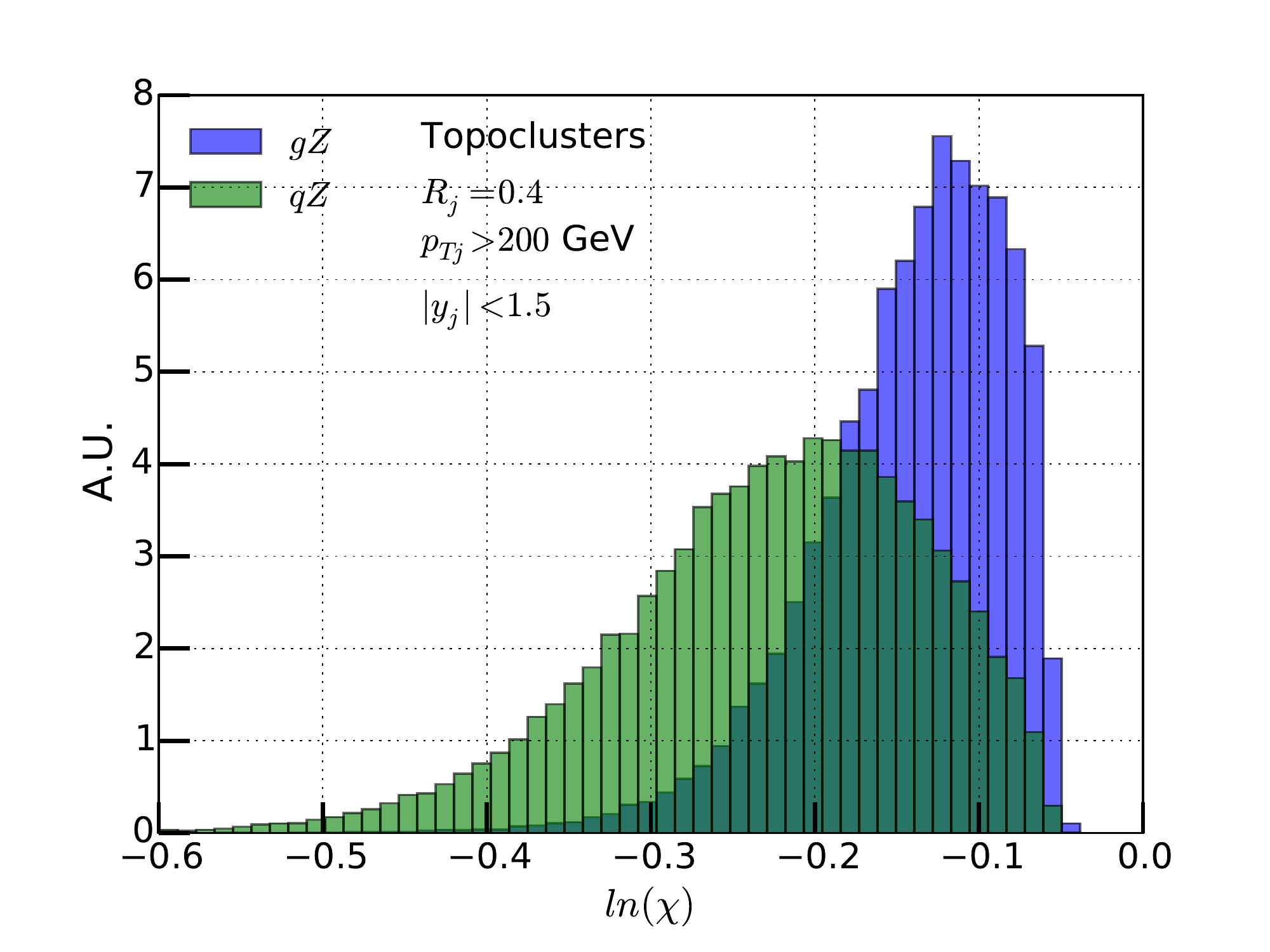}
\caption{ Distributions of $r_2$ (left) and $\ln(\chi)$ (right) in $Z + \mathrm{jet}$ events. The leading jet with $|y_j|<1.5$ is reconstructed from massive topoclusters.}
\label{fig:DIST_xZ_y15_GvQ_SD}
\end{figure}

\subsubsection{Energy correlation functions}
\label{sec:ecf}

We now turn to an established family of observables with the potential to distinguish between quark and gluon jets: energy correlation functions and ratios derived from these functions \cite{Larkoski:2013eya} \cite{Larkoski:2014gra}. The energy correlation functions are defined by
\begin{equation}
\label{eq:ecf}
\begin{split}
&ECF(0,\beta) = 1, \\
&ECF(1,\beta) = \sum\limits_{i\in J} p_{T,i},\\
&ECF(2,\beta) = \sum\limits_{i<j\in J} p_{T,i}p_{T,j}\left(R_{ij}\right)^\beta, \\
&ECF(N,\beta) = \sum\limits_{i_1<i_2<..<i_n\in J} \left(\prod\limits_{a=1}^{N}p_{T,i_a}\right)\left(\prod\limits_{b=1}^{N-1}\prod\limits_{c=b+1}^{N}R_{i_bi_c}\right)^\beta, \\
\end{split}
\end{equation}
From these, we can define the ratios
\begin{equation}
\label{eq:ecfr}
\begin{split}
&r_N^{(\beta)} = \frac{ECF(N+1,\beta)}{ECF(N,\beta)}, \\
&C_N^{(\beta)} = \frac{r_N^{(\beta)}}{r_{N-1}^{(\beta)}} =  \frac{ECF(N+1,\beta)ECF(N-1,\beta)}{ECF(N,\beta)^2}. \\
\end{split}
\end{equation}
The sums run over the constituents $i$ of the jet $J$. We tested several jet shapes from this family ($r_0$, $r_1$, $r_2$, $C_1$, $C_2$). We also examined the variable $D_2$, defined in \cite{Larkoski:2014gra}, and N-subjettiness variables \cite{Thaler:2010tr} ($\tau_1$, $\tau_2$, $\tau_2/\tau_1$, $\tau_3/\tau_2$) with the angular exponent in all cases set to $\beta = 0.2$ for quark/gluon tagging, as suggested by the authors. Of those, $C_1$, $r_1$, and $r_2$ provided the best background rejection. If we express $C_1$, and $r_2$ explicitly using equations \ref{eq:ecf} and \ref{eq:ecfr} we find

\begin{equation}
\label{eq:ecfvars}
\begin{split}
&C_1 = \frac{\sum\limits_{i<j\in J} p_{T,i}p_{T,j}\left(R_{ij}\right)^{0.2}}{\sum\limits_{i,j\in J} p_{T,i}p_{T,j}},\\
&r_2 = \frac{\sum\limits_{i<j<k\in J} p_{T,i}p_{T,j}p_{T,k}\left(R_{ij}R_{ik}R_{kj}\right)^{0.2}}{\sum\limits_{i<j\in J} p_{T,i}p_{T,j}\left(R_{ij}\right)^{0.2}}. 
\end{split}
\end{equation}

It is evident that the numerator of $C_1$ is larger if the radiation within the jet is split evenly between two or more distinct directions than if most of the energy is clustered within a small angular area. Therefore, $C_1$ is differentiates between 1-prong and 2-prong jets. The variable $r_2$ is larger if the radiation is localised in three directions and smaller for 2-prong and 1-prong jets. 

The justification for the relatively small angular exponent comes from eq.\ (3.22) in \cite{Larkoski:2013eya}. The authors find a power law relation between the cumulative distributions of the $C_1$ variable for gluon and quark jets. A small $\beta$ increases the magnitude of the power that relates the two distributions, thereby directly contributing to a better ROC curve. Note, however, that perturbative splitting probabilities have singularities at $R_{ij} = 0$. Thus the positive powers of $R_{ij}$ are needed to keep the observables from being infrared unsafe against collinear splittings. With a power $\beta = 0.2$, our observables are technically infrared safe, but they are quite sensitive to infrared effects.

As a result of the asymmetry in the quark and gluon-jet distributions in figure~\ref{fig:DIST_xZ_y15_GvQ_SD}, we find a different ROC curve for quark compared to gluon tagging\footnote{According to eq.\ (3.7) in \cite{Larkoski:2013eya}, 
if we were to perform quark tagging using $C_1$, the background fake rate as a function of the signal efficiency would be given by 
\begin{equation}
\varepsilon_b(\varepsilon_s)=\varepsilon_s^{C_A/C_F}=\varepsilon_s^{2.25}\;.
\end{equation}
Thus the gluon fake rate at 50\% quark efficiency is $\varepsilon_b(0.5)\approx 0.21$. If we were to do the opposite and tag gluon jets at the expense of quark jets, then we would have to make the cut in the opposite direction of the $C_1$ distribution. Using the same relation between quark and gluon acceptances, we conclude that, when we retain 50\% of the gluon jets in a sample, the fake rate from quark jets is $1-(1-0.5)^{\frac{1}{2.25}}\approx 0.27$. Therefore, the same discriminating variable can perform differently depending on the type of tagging we would like to do. This asymmetry is strongly in favour of quark tagging for all of the variables that we study, as will become evident in the following sections.}. 
For example, if we want to tag a quark and impose a cut on $\ln(\chi(q,g)) > 0.3$, we achieve $\varepsilon_s\simeq 0.21$ and $\varepsilon_b \simeq 0.017$. If we instead tag a gluon by requiring $\ln(\chi(g,q))$ to be bigger than a specific value, for $\varepsilon_s\simeq 0.21$ we find only $\varepsilon_b \simeq 0.05$.

\begin{figure} 
\centering
\includegraphics[width=0.49\linewidth]{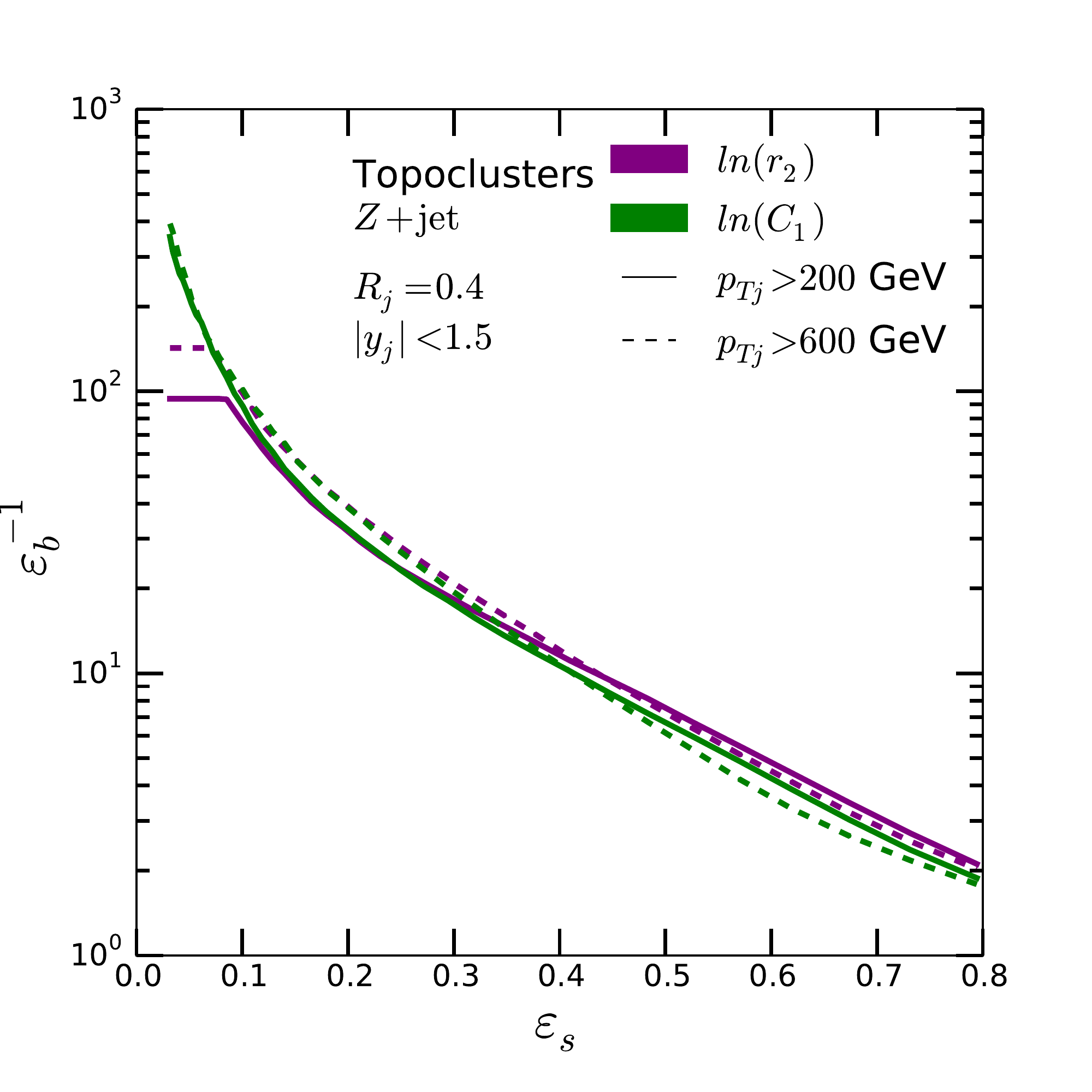}
\includegraphics[width=0.49\linewidth]{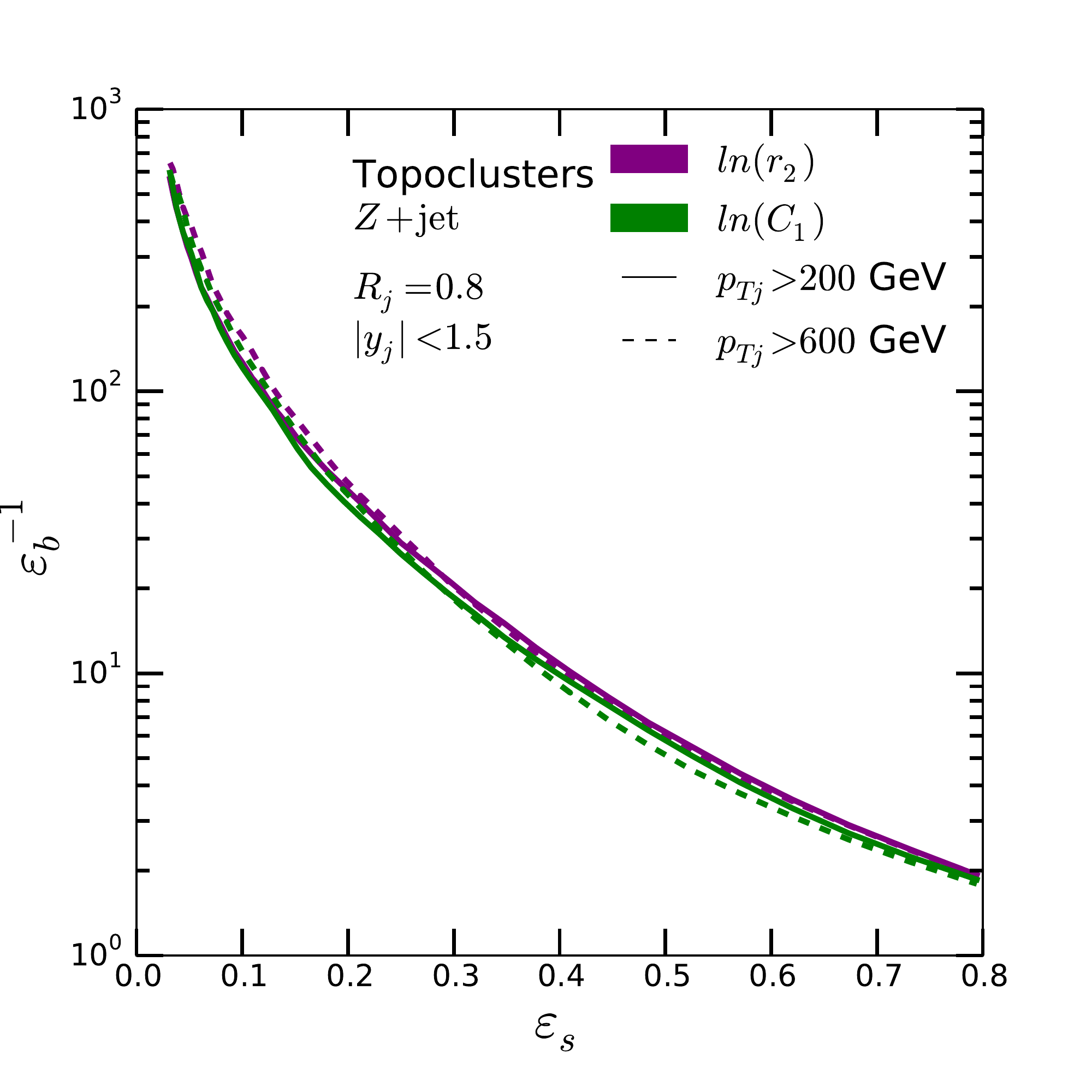}
\includegraphics[width=0.49\linewidth]{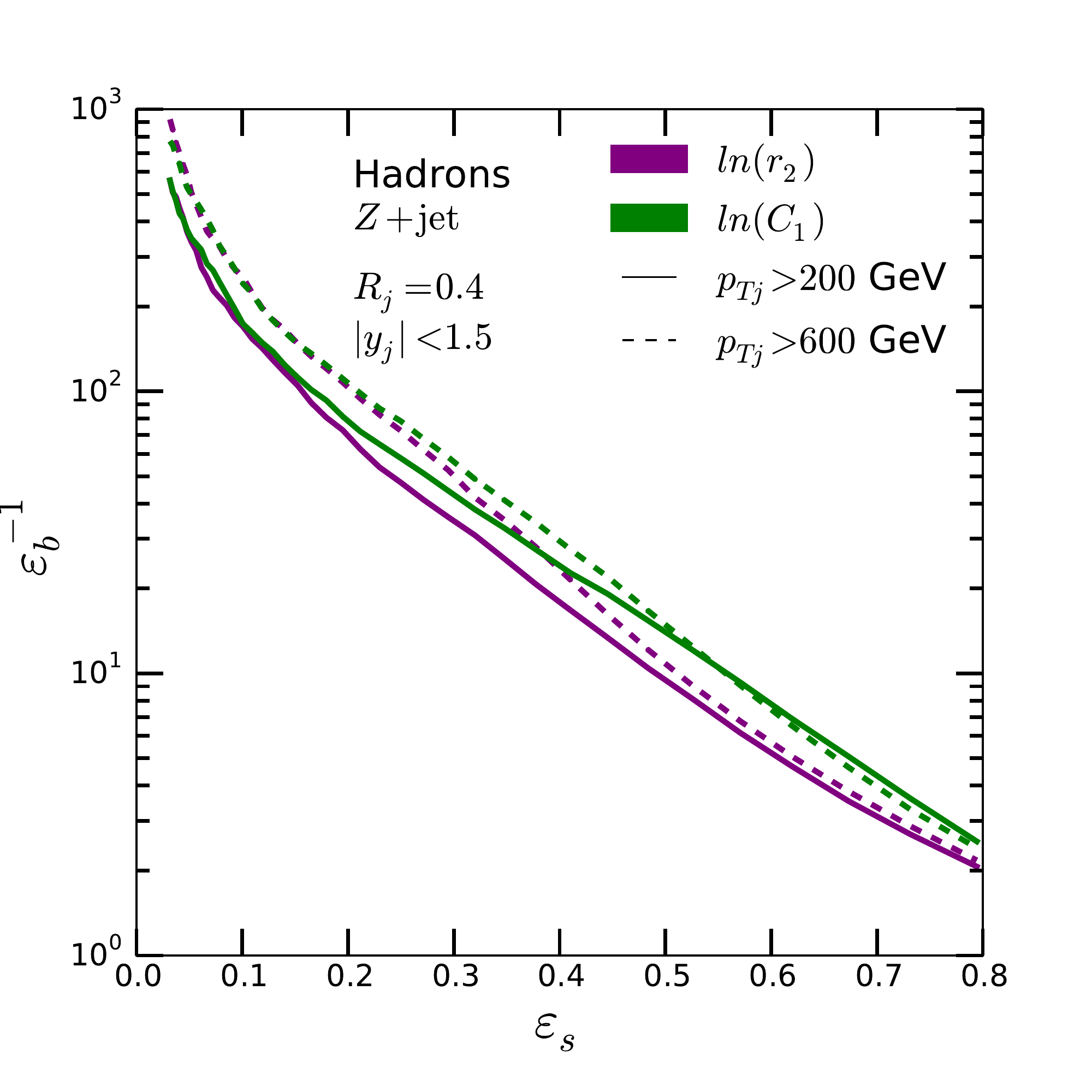}
\includegraphics[width=0.49\linewidth]{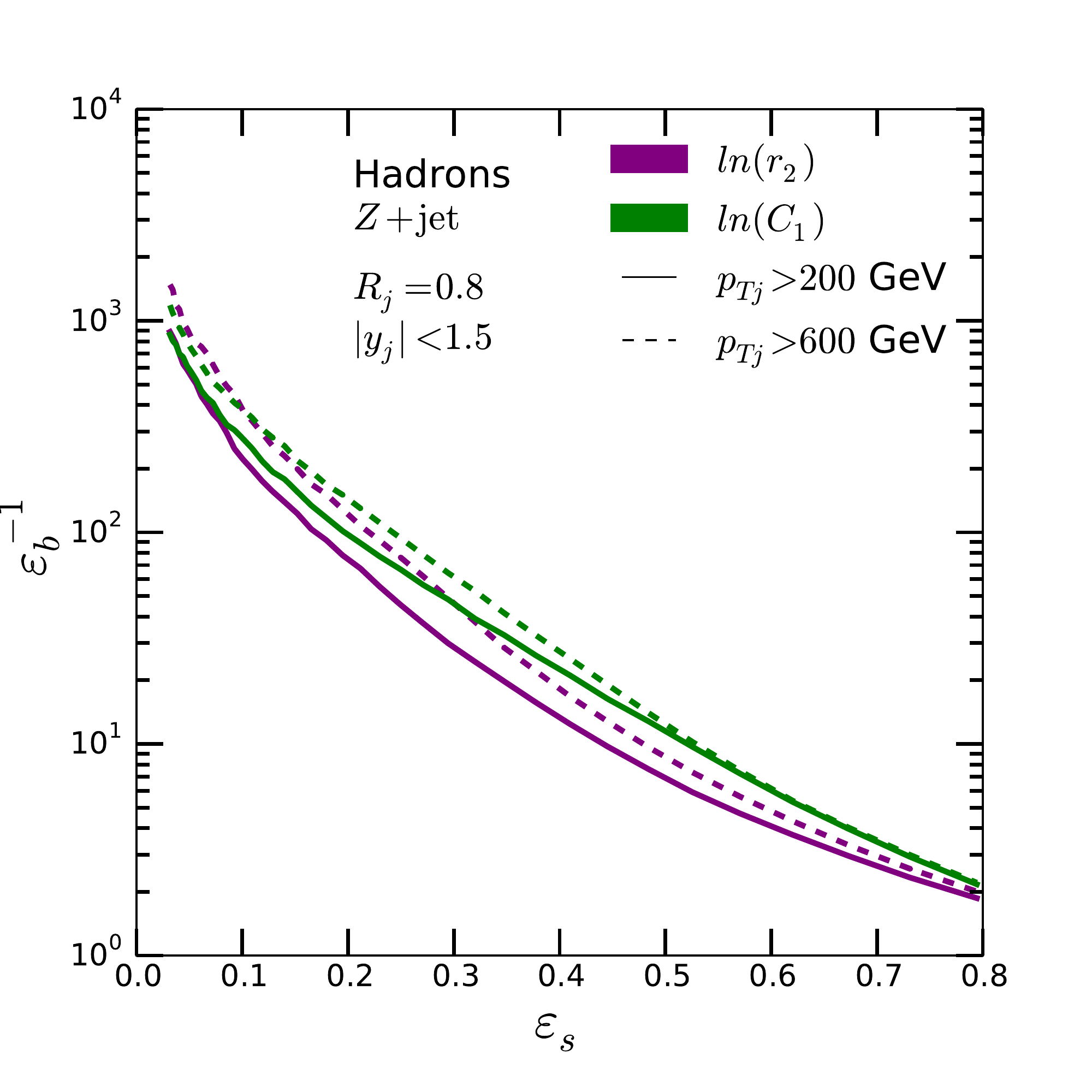}
\caption{ROC plots comparing $r_2$ and $C_1$ performance at different jet $p_T$. The top row uses massive topoclusters as inputs and the bottom uses hadrons. The left (right) column uses jets with small (large) radius.}
\label{fig:ROC_xZ_y15_R2vC1_Rfixed}
\end{figure}

\begin{figure} 
\centering
\includegraphics[width=0.49\linewidth]{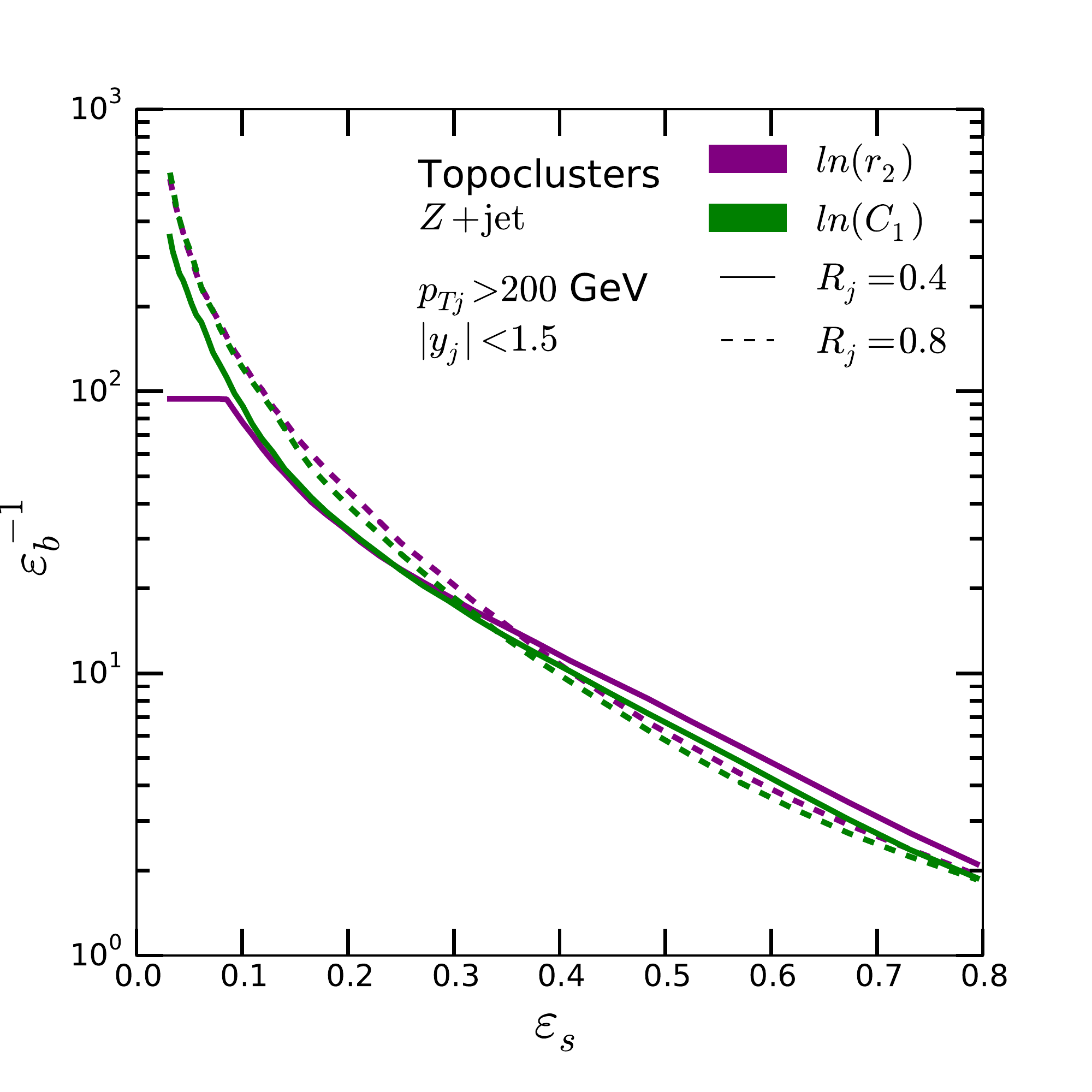}
\includegraphics[width=0.49\linewidth]{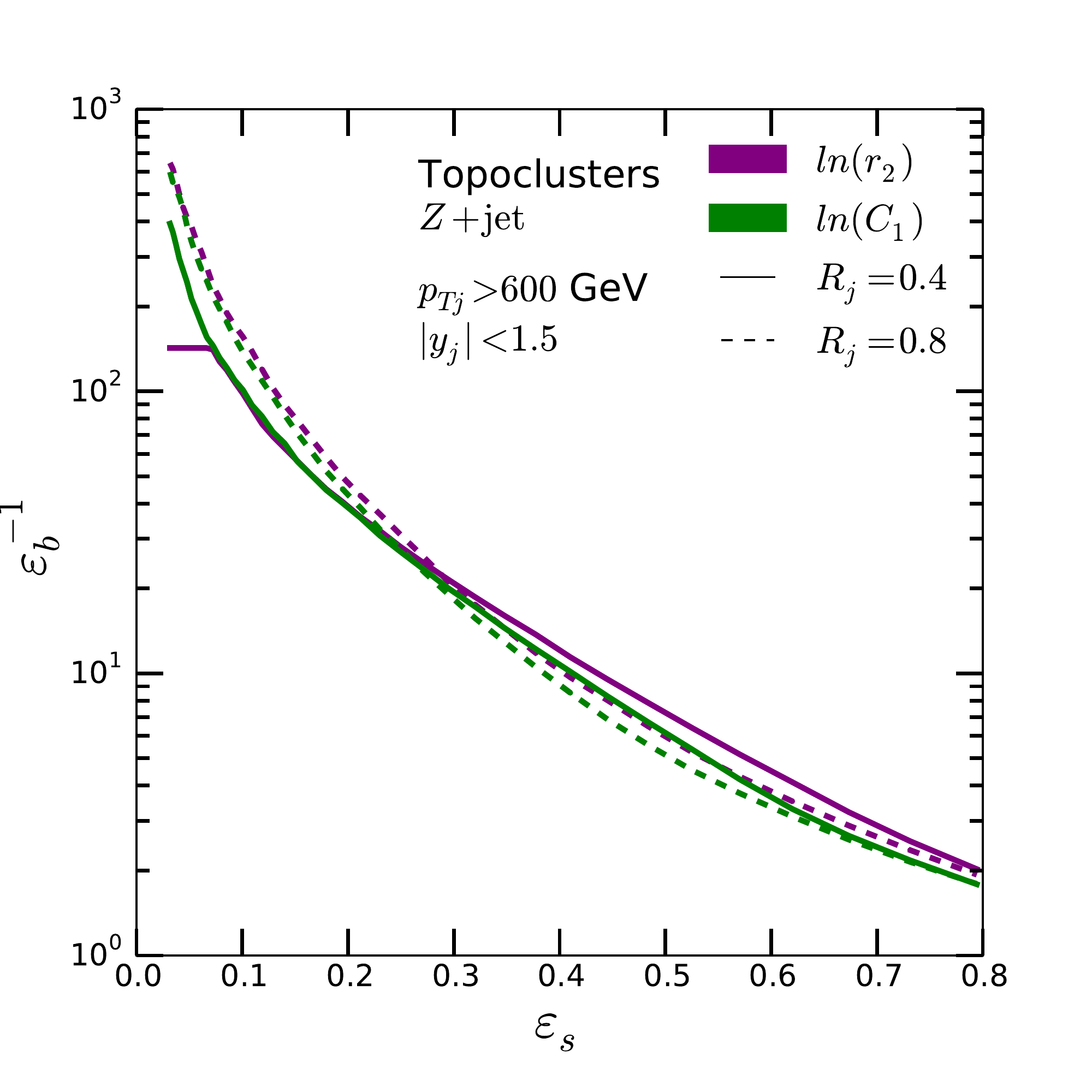}
\includegraphics[width=0.49\linewidth]{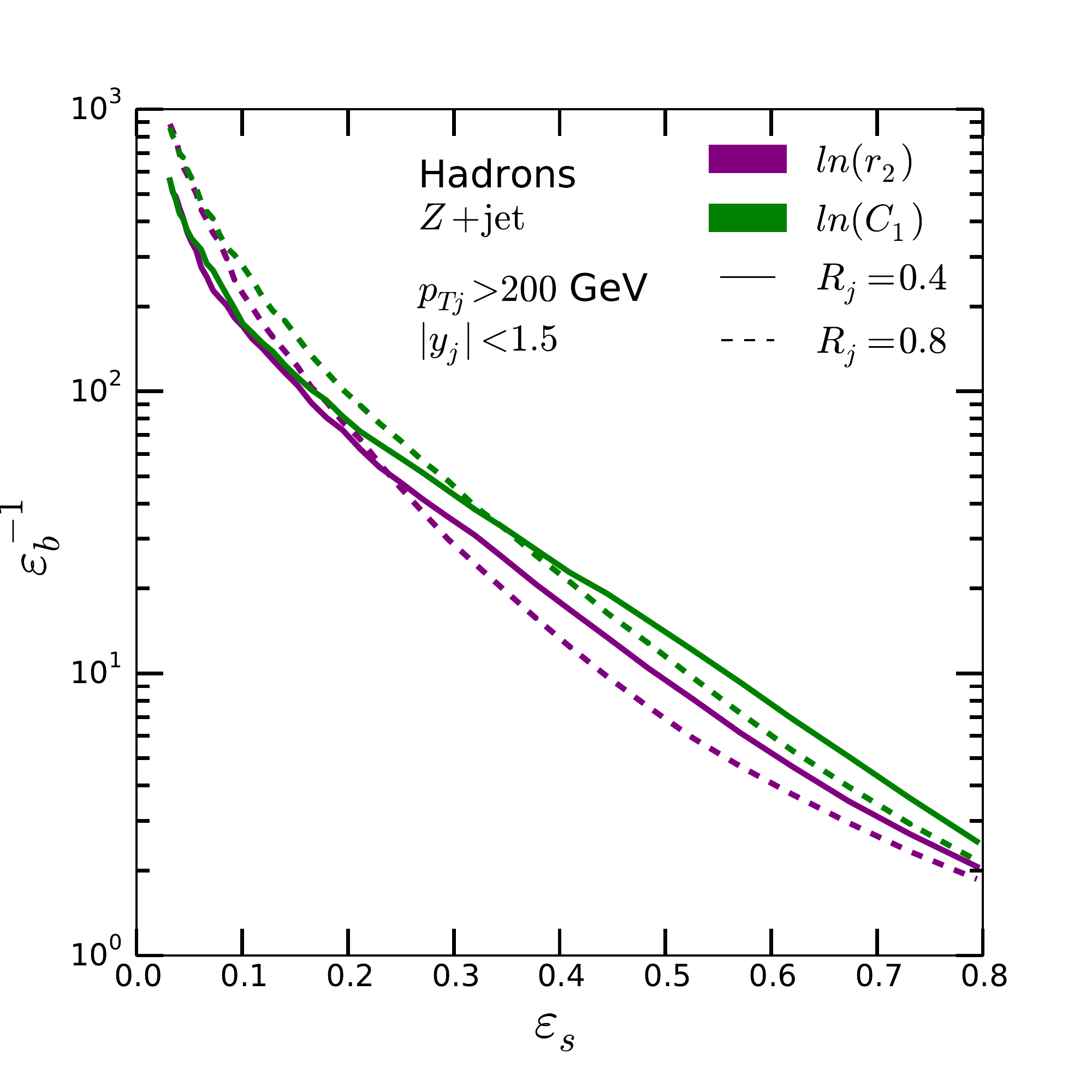}
\includegraphics[width=0.49\linewidth]{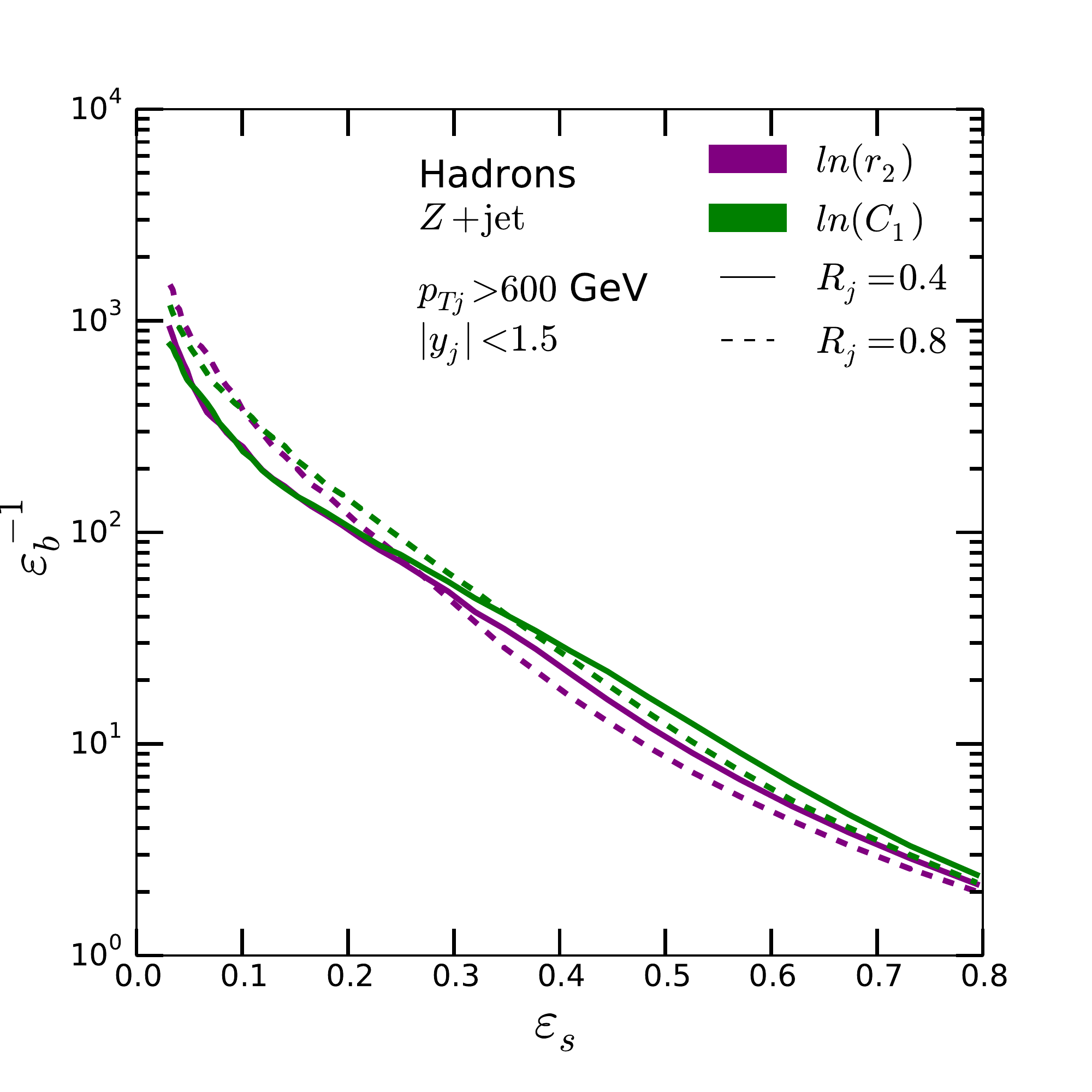}
\caption{ ROC plots comparing $r_2$ and $C_1$ performance at different jet radii. The top row uses massive topoclusters as inputs and the bottom uses hadrons. The left (right) column uses jets with small (large) boost.}
\label{fig:ROC_xZ_y15_R2vC1_pTfixed}
\end{figure}

A preliminary study of quark tagging with energy correlation variables uncovers some trends. As expected from the discussion in \cite{Larkoski:2013eya}, we find that the variable $C_1$ is favored over $r_2$ over a large variety of jet parameters as long as the jets are reconstructed from hadrons. This can be seen in the bottom rows of figures \ref{fig:ROC_xZ_y15_R2vC1_Rfixed} and \ref{fig:ROC_xZ_y15_R2vC1_pTfixed}, where its background fake rate is about 70\% to 60\% of that obtained with $r_2$ at moderate signal efficiency. This difference diminishes at small signal efficiency. A common trend among the energy correlation variables is that increasing the radius of the jet reduces the performance at moderate and large $\varepsilon_s$, but leads to improvement at low signal efficiency. This effect is true for any jet type as can be seen in the four plots of figure \ref{fig:ROC_xZ_y15_R2vC1_pTfixed}. Another trend in figure \ref{fig:ROC_xZ_y15_R2vC1_Rfixed} is that for jets built from hadron inputs, a larger $p_T$ limit increasingly improves background rejection as the signal cut becomes more stringent. This effect does not translate to topocluster inputs where the discrimination of the energy correlation variables remains largely independent of the jet's transverse momentum. 


\section{Comparisons of tagging results}
\label{sec:tag}

\begin{figure}[!h]
\centering 
\includegraphics[width=0.4\linewidth]{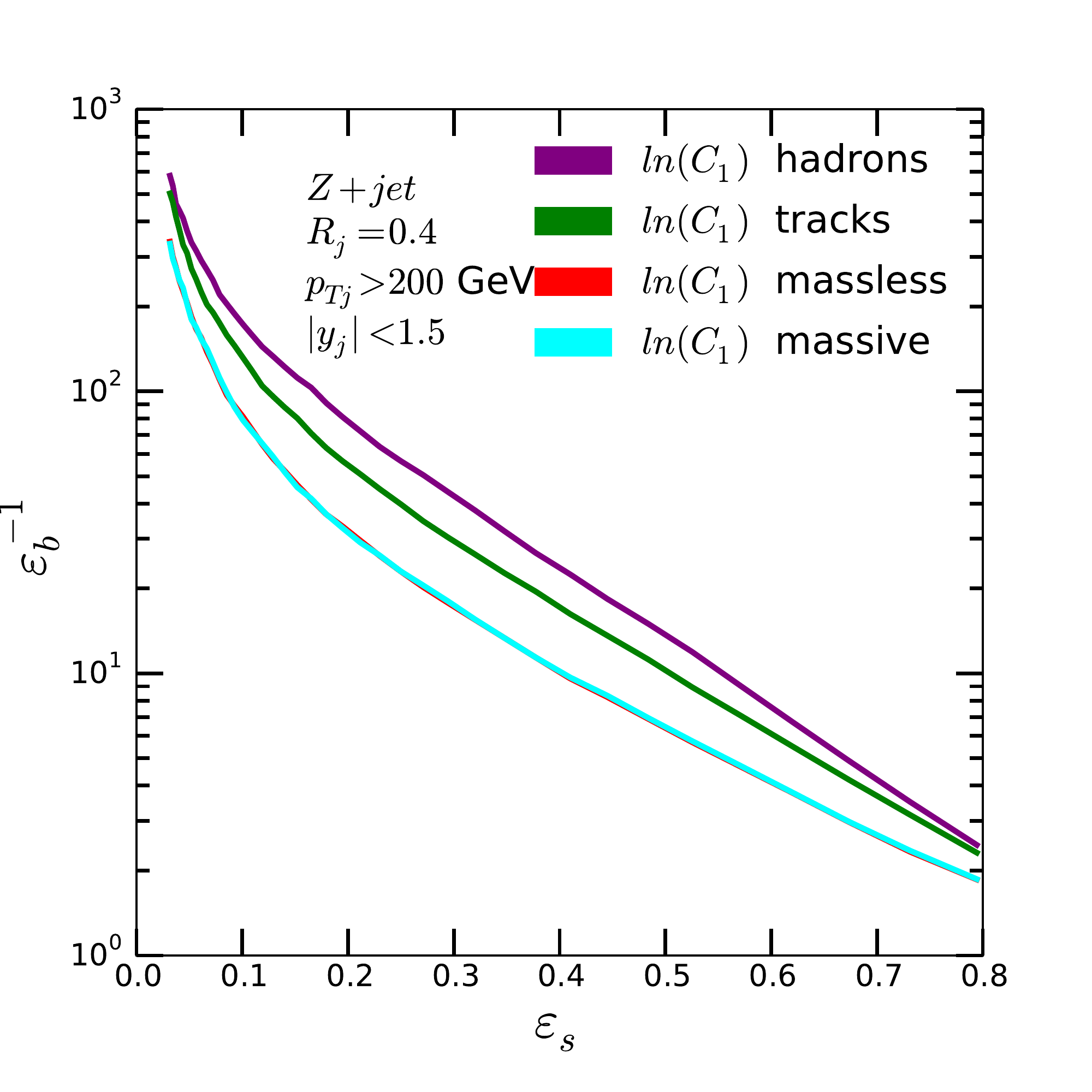}
\includegraphics[width=0.4\linewidth]{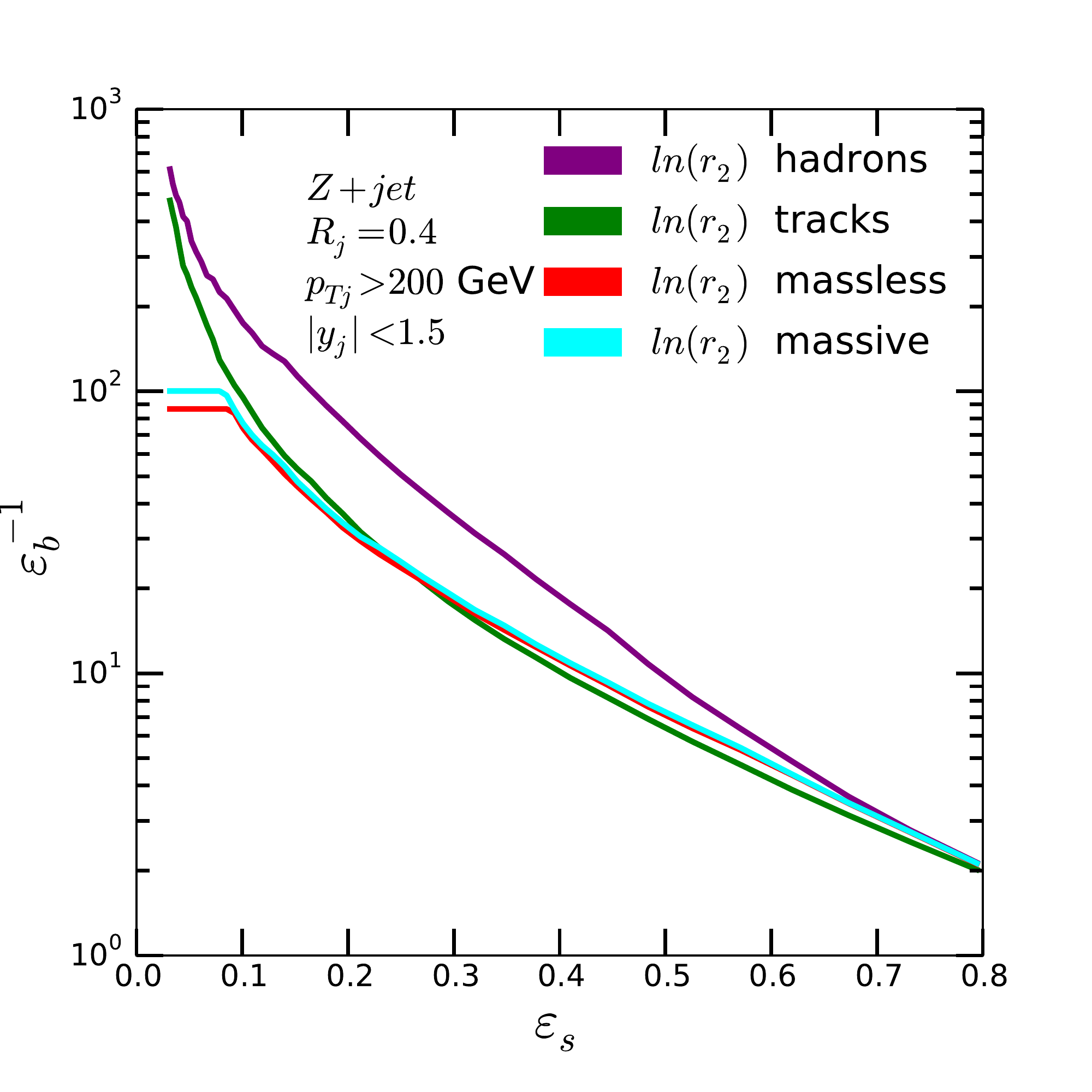}\\
\includegraphics[width=0.4\linewidth]{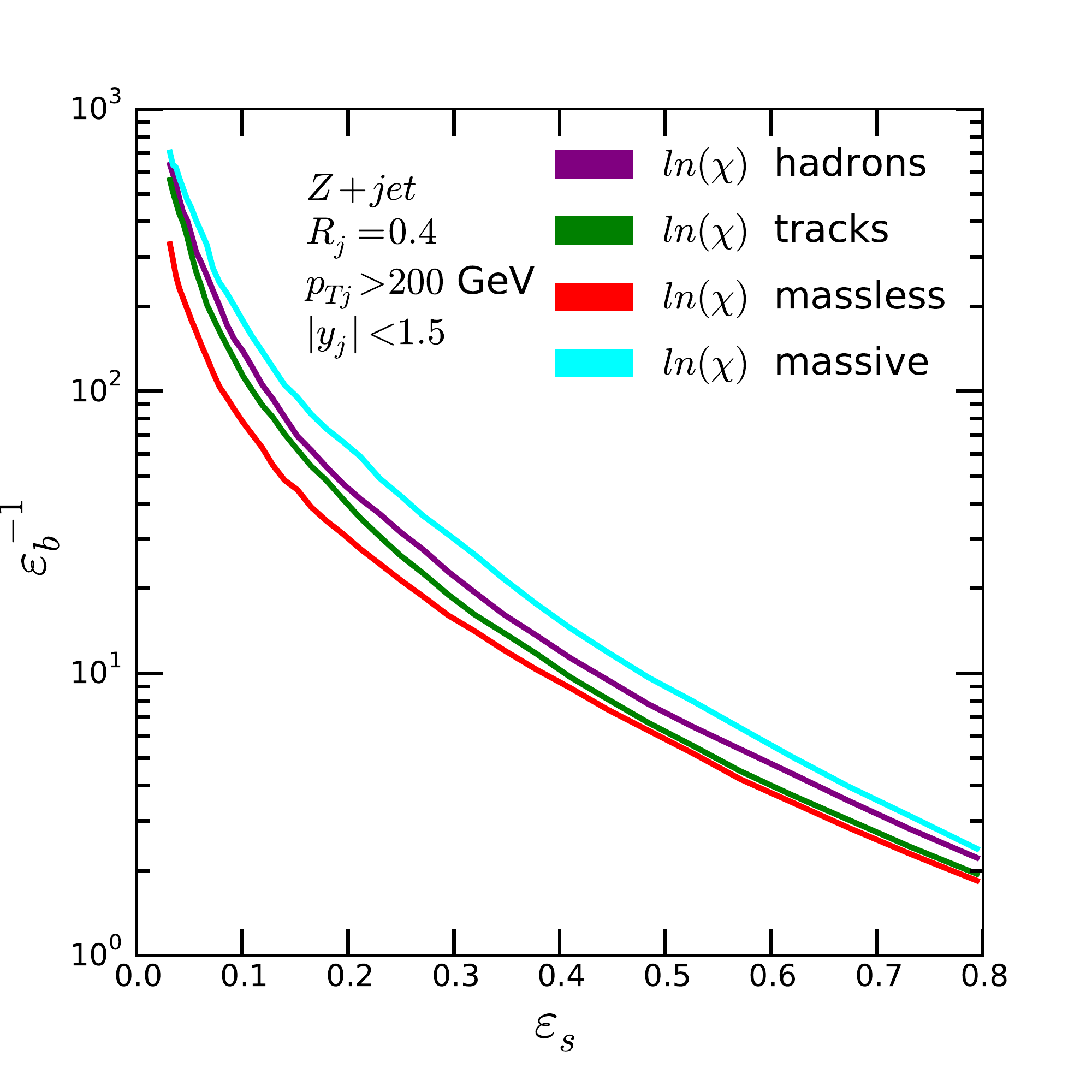}
\includegraphics[width=0.4\linewidth]{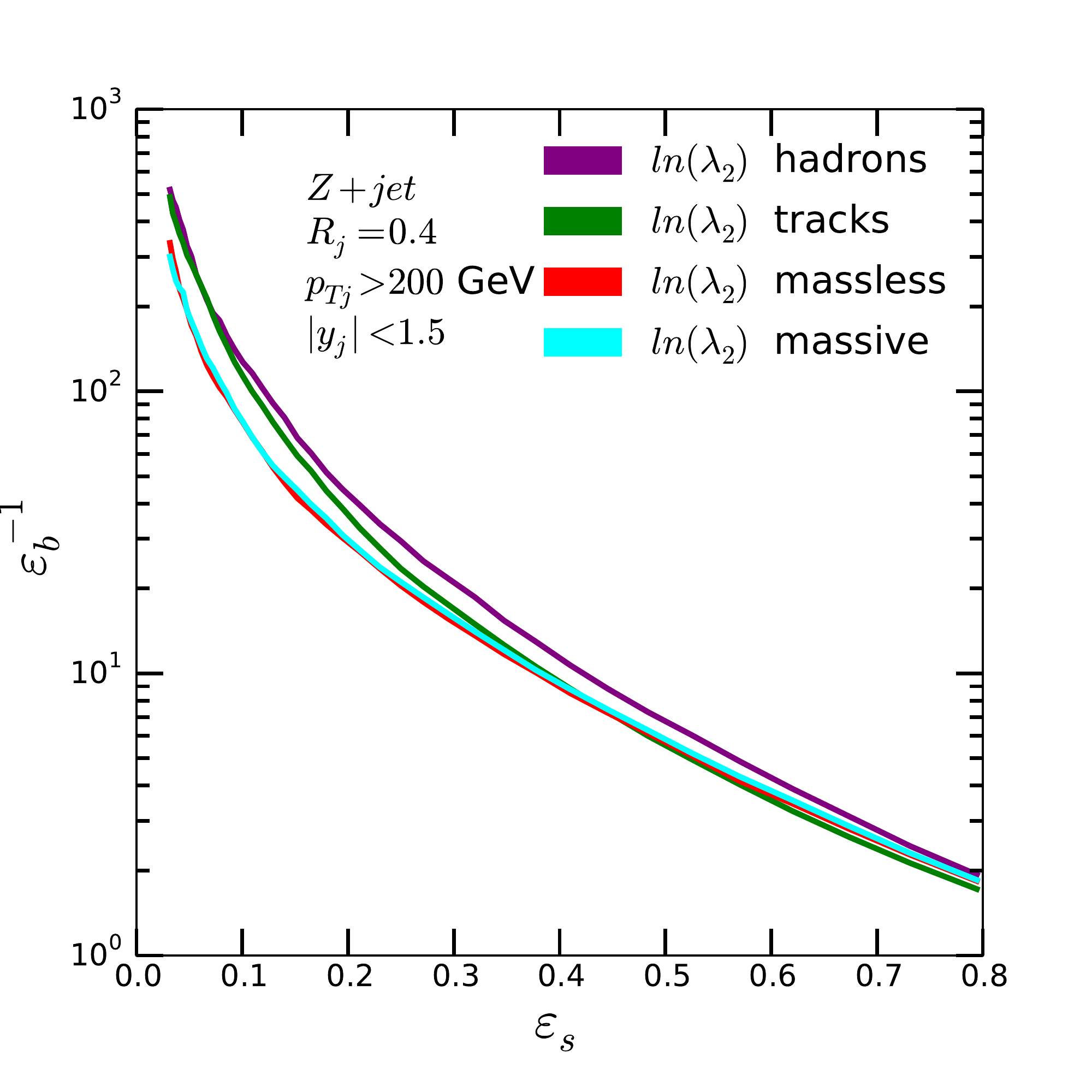}

\caption{ROC curves of the leading jet with $|y|<1.5$ for $C_1$ (upper left), $r_2$ (upper right), $\chi$ (lower left), $\lambda_2$ (lower right) and using hadrons, charged tracks, massless and massive topoclusters as inputs.}
\label{fig:ROC_AllVarsAllInputs}
\end{figure}

In this section, we compare methods for distinguishing quark jets from gluon jets. 

We begin in figure~\ref{fig:ROC_AllVarsAllInputs} with a study of the dependence of four observables on the choice of input objects: hadrons, tracks, massless topoclusters, and massive topoclusters. In each panel of figure~\ref{fig:ROC_AllVarsAllInputs}, we show the dependence on input objects for one observable, $C_1$, $r_2$, $\chi$ from shower deconstruction with a single microjet, and the angularity variable $\lambda_2$ \cite{Larkoski:2014pca} defined by
\begin{equation}
\lambda_2 = \sum_{i\in J} p_{T,i}\, \theta_i^2 \Bigg/ \left(\sum_{i\in J} p_{T,i}\right)
\;.
\end{equation}
If the input constituents $i$ are massless, $\lambda_2$ is approximately $2 M_J^2/p_{T,J}^2$, where $M_J$ is the jet mass. We show $\lambda_2$ because it is rather similar to $\chi$ if the input objects are all massless. However, $\chi$ is sensitive to the masses of the input objects while $\lambda_2$ is not. The ROC curves we show are obtained from distributions like the ones in figure~\ref{fig:DIST_xZ_y15_GvQ_SD} by swiping a cut from one end to the other. 

All variables show some dependence on the input objects. Hadrons give the best results for $C_1$ and $r_2$, although detecting neutral as well as charged hadrons is not as realistic as the other input choices. After that, $C_1$ does best with tracks, while all of the other input choices work equally well for $r_2$. The variable $\lambda_2$ gives results that are rather insensitive to the choice of inputs, and not sensitive at all to the choice between massive and massless topoclusters. In contrast, the results for $\chi$ are significantly better with massive topocluster inputs than with massless topocluster inputs. This is to be expected because the topocluster mass $\mu_J$ is one of the variables used in the calculation of $\chi$ in eq.~(\ref{eq:SD1}). With massless topoclusters as input, we are forced to set $\mu_J$ to a minimum value, $\mu_J = 1 \GeV$, but this loses information. Perhaps surprisingly, $\chi$ works better with massive topocluster inputs than with all hadrons as inputs. This is because our definition of massive topoclusters drops topoclusters with $p_\LT < 1 \GeV$, on the grounds that such topoclusters would be experimentally unobservable. Dropping these low $p_\LT$ topoclusters also helps to suppress unwanted contributions from initial state radiation, making $\chi$ more sensitive to the distinguishing features of quark jets compared to gluon jets. \

\begin{figure}[!h]
\centering 
\includegraphics[width=0.5\linewidth]{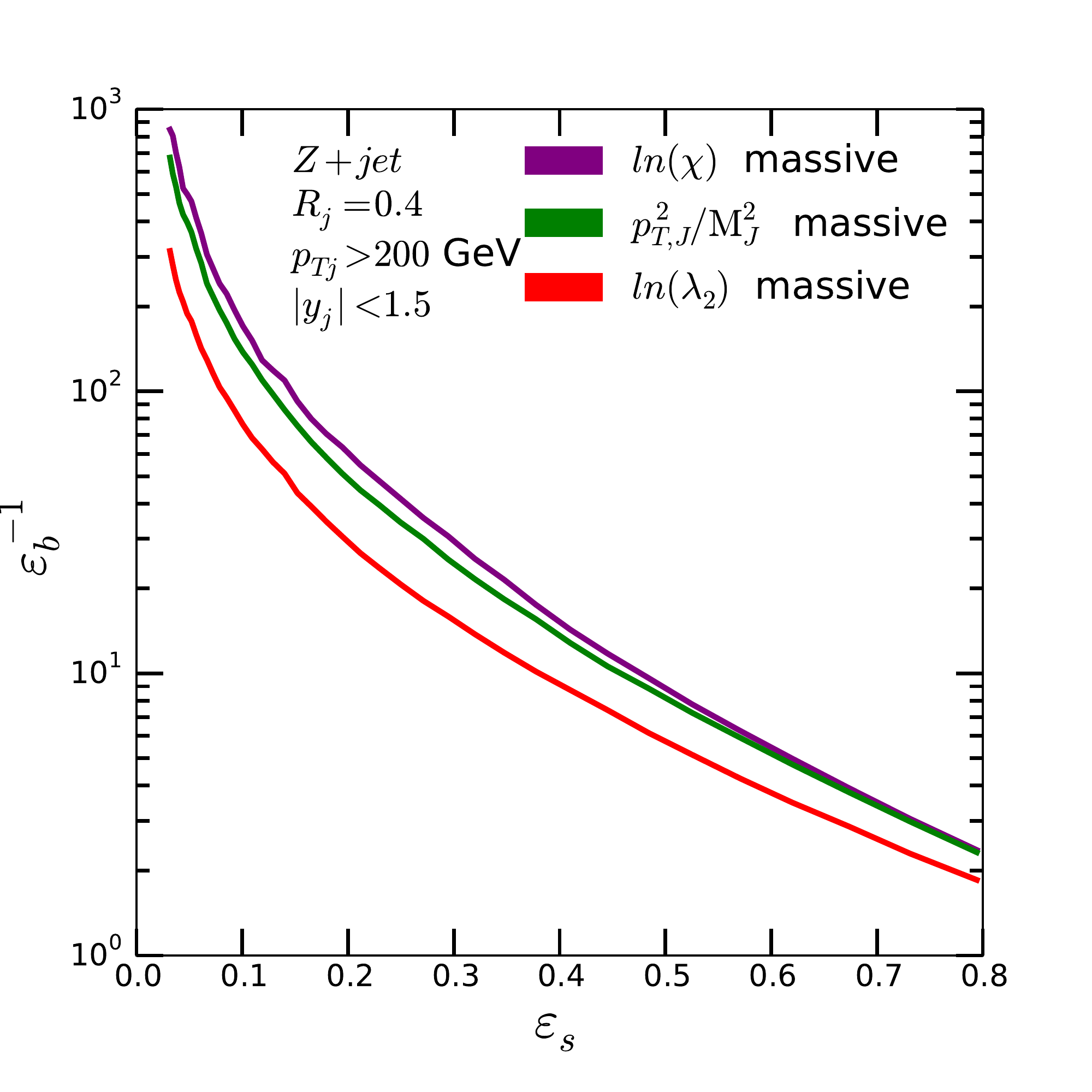}
\caption{ROC curves of the leading jet with $|y|<1.5$. We compare $\chi$ to $\lambda_2$ and a simple squared ratio of the jet transverse momentum and mass using massive topoclusters as inputs.}
\label{fig:ROC_SDvL2vRatio}
\end{figure}

We compare directly $\lambda_2$ to $\chi$ in figure \ref{fig:ROC_SDvL2vRatio}. It is evident that shower deconstruction with massive topoclusters is better than the angularity variable. The latter is equivalent to the squared ratio between the jet mass and $p_T$ as long as the input objects are massless and nearly collinear. The former condition is not satisfied in our case; therefore, we add the explicit ratio as a separate variable in the plot. Although much better than $\lambda_2$ it still performs worse than shower deconstruction.

\begin{figure}[!h]
\centering 
\includegraphics[width=1.1\linewidth]{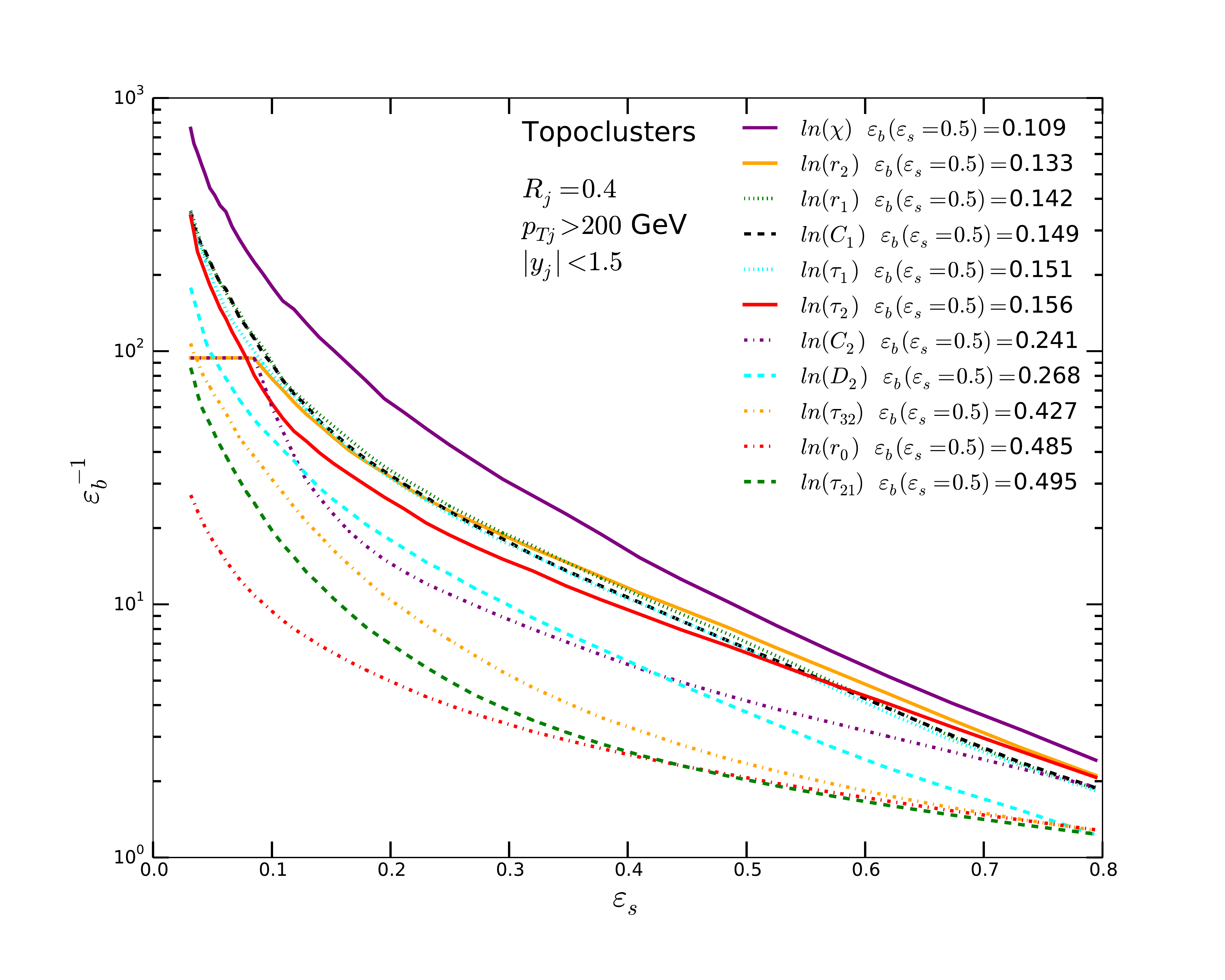}
\caption{ ROC curves for all distributions for quark tagging of $Z + \mathrm{jet}$ events. Leading jet with $|y|<1.5$ reconstructed from massive topoclusters.}
\label{fig:ROC_AllVars}
\end{figure}

We turn next to a comparison of several observables that can be used for quark-gluon discrimination. Here, and in the studies that follow, we use massive topocluster inputs. The ROC curves for the observables are shown in figure~\ref{fig:ROC_AllVars}. For shower deconstruction, we use just one microjet equal to the whole fat jet. Shower deconstruction $\chi$ has the best ROC curve. However, there is no dominant jet-shape or energy correlation function variable. Instead, there is a tier of closely spaced ROC curves. The top tier contains [$r_2$, $r_1$, $C_1$, $\tau_1$, $\tau_2$] and spreads within a band of about $\Delta\varepsilon_b \approx 20\%$ across the entire $\varepsilon_s$ range. The ratio $r_2$ consistently performs better at moderate and large signal efficiency and remains competitive at small efficiency. Therefore, to the benefit of clarity of the results we are going to present, we believe it is acceptable to compare our choice of $\chi$ with $r_2$.

\begin{figure} 
\centering
\includegraphics[width=0.49\linewidth]{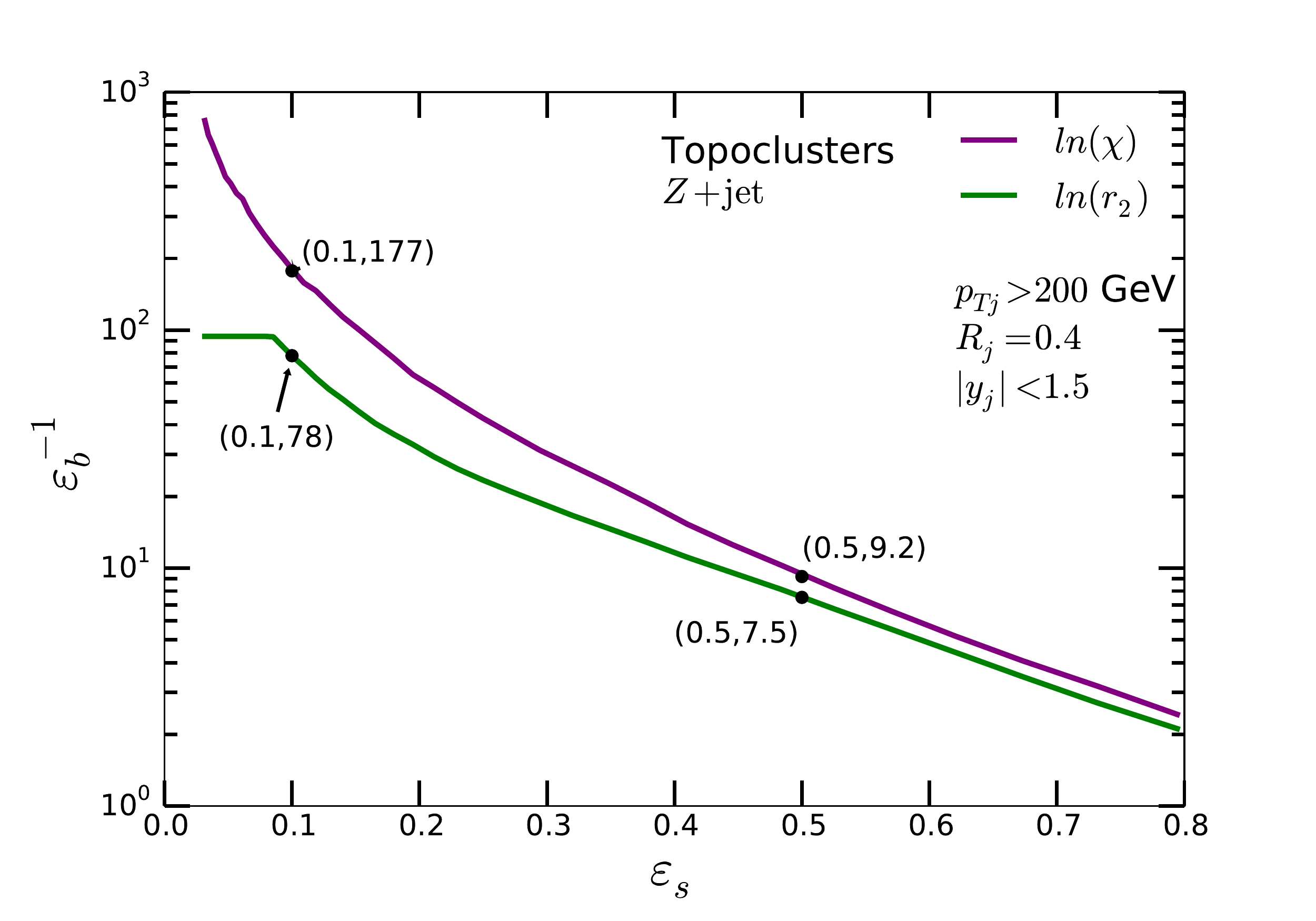}
\includegraphics[width=0.49\linewidth]{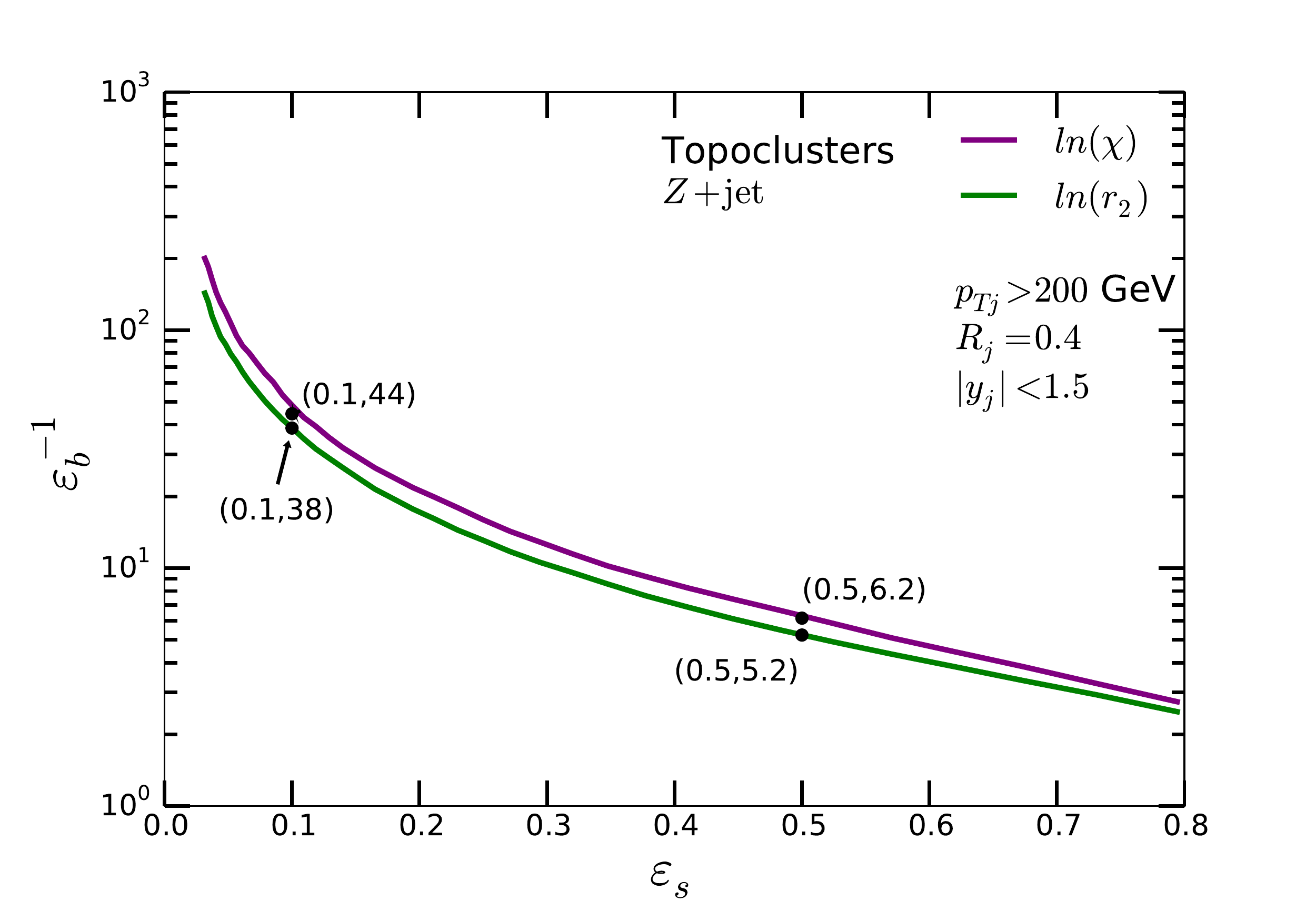}
\caption{Left: ROC curves for quark tagging and gluon rejection from $Z + \mathrm{jet}$ events. Right: ROC curves for gluon tagging and quark rejection from $Z + \mathrm{jet}$ events. The leading jet with $|y|<1.5$ is reconstructed from massive topoclusters.}
\label{fig:ROC_xZ_y15_R04_pT200_T_SDvR2}
\end{figure}

In figure~\ref{fig:ROC_xZ_y15_R04_pT200_T_SDvR2} we show the ROC curves for the observables $\chi$ and $r_2$ for quark tagging (left) and gluon tagging (right) respectively. It is immediately apparent that quark tagging performs much better than gluon tagging, as already suggested by the analytic approximation of \cite{Larkoski:2013eya} and the discussion in section~\ref{sec:ecf}. At small efficiencies the gluon rejection in the left plot is four times better than the quark rejection on the right for shower deconstruction and two times better for $r_2$.  One might anticipate this trend by looking at the probability densities of the variables. It is true for both observables, although more obvious for $\chi$, that the quark distribution drops off slower at the gluon-like region end (large values) than the gluon distribution at the quark-like end (low values). This asymmetry allows for the substantial gluon rejection at small quark efficiency. Another feature is that the single-branch $\chi$ performs better than $r_2$ across the entire signal efficiency range in both quark and gluon tagging. For quark tagging it is about 20\% better at moderate efficiencies and about a factor of two better at low efficiency. The difference is notably smaller when we attempt gluon tagging and almost disappears at low efficiency if we replace $r_2$ with a better performing energy correlation variable at that efficiency region. An obvious feature, although in a region that we do not explore, in the $r_2$ ROC curve is the plateau at $\varepsilon_s<0.1$. It is an artefact from binning of jets on which the variable cannot be defined. The ratio $r_2$ needs at least 3 jet constituents. The condition is not always met with $R=0.4$ jets reconstructed from topoclusters. More careful treatment of this bin can remove the plateau. It has to be noted that the energy correlation and N-subjettiness variables are used without optimisation with the recommended value $\beta = 0.2$ for quark and gluon tagging. Hence, there might be room for further improvements.

\begin{figure}[!h]
\centering 
\includegraphics[width=1.1\linewidth]{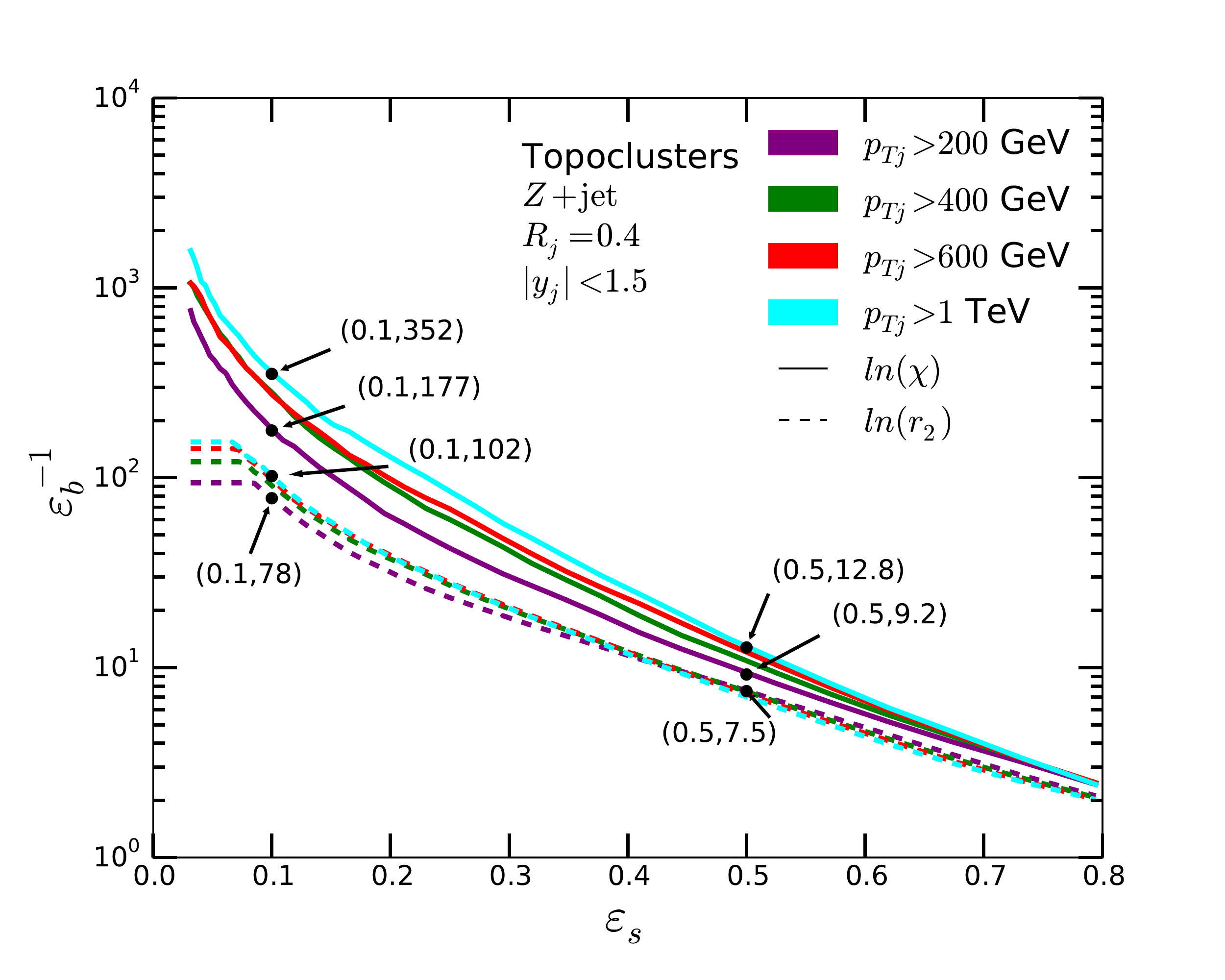}
\caption{ ROC curves for all $p_T$ bins for quark tagging of $Z + \mathrm{jet}$ events with $\chi$ and $r_2$. The leading jet with $|y|<1.5$ is reconstructed from massive topoclusters. The solid lines correspond to $\ln(\chi)$ of shower deconstruction and the dashed lines to the energy correlation function $\ln(r_2)$.}
\label{fig:ROC_xZ_y15_QvG_Rfixed_SDvR2}
\end{figure}

The results in figure~\ref{fig:ROC_xZ_y15_R04_pT200_T_SDvR2} are obtained from jets with $p_T>200\,\mr{GeV}$. Collisions at the LHC can provide sufficient energy for much more boosted jets, either from a heavy particle decay or from a recoil in a high $p_T$ event. In figure~\ref{fig:ROC_xZ_y15_QvG_Rfixed_SDvR2} we see the effect on quark tagging from increasing the jet transverse momentum. While we saw in figure \ref{fig:ROC_xZ_y15_R2vC1_Rfixed} that increasing the jet $p_T$ beyond $200\,\mr{GeV}$ has little or no effect on energy correlation variables, there is a distinct improvement in quark tagging with shower deconstruction as the jet gets more boosted. Moreover, the improvement is significant at 50\% signal efficiency (40\% better background rejection) and it steadily widens the difference between the $\chi$ and $r_2$ performance, leading to a factor of three better gluon rejection by $\chi$ than $r_2$ at $\varepsilon_s=0.1$. 

\begin{figure}[!t]
\centering
\includegraphics[width=0.49\linewidth]{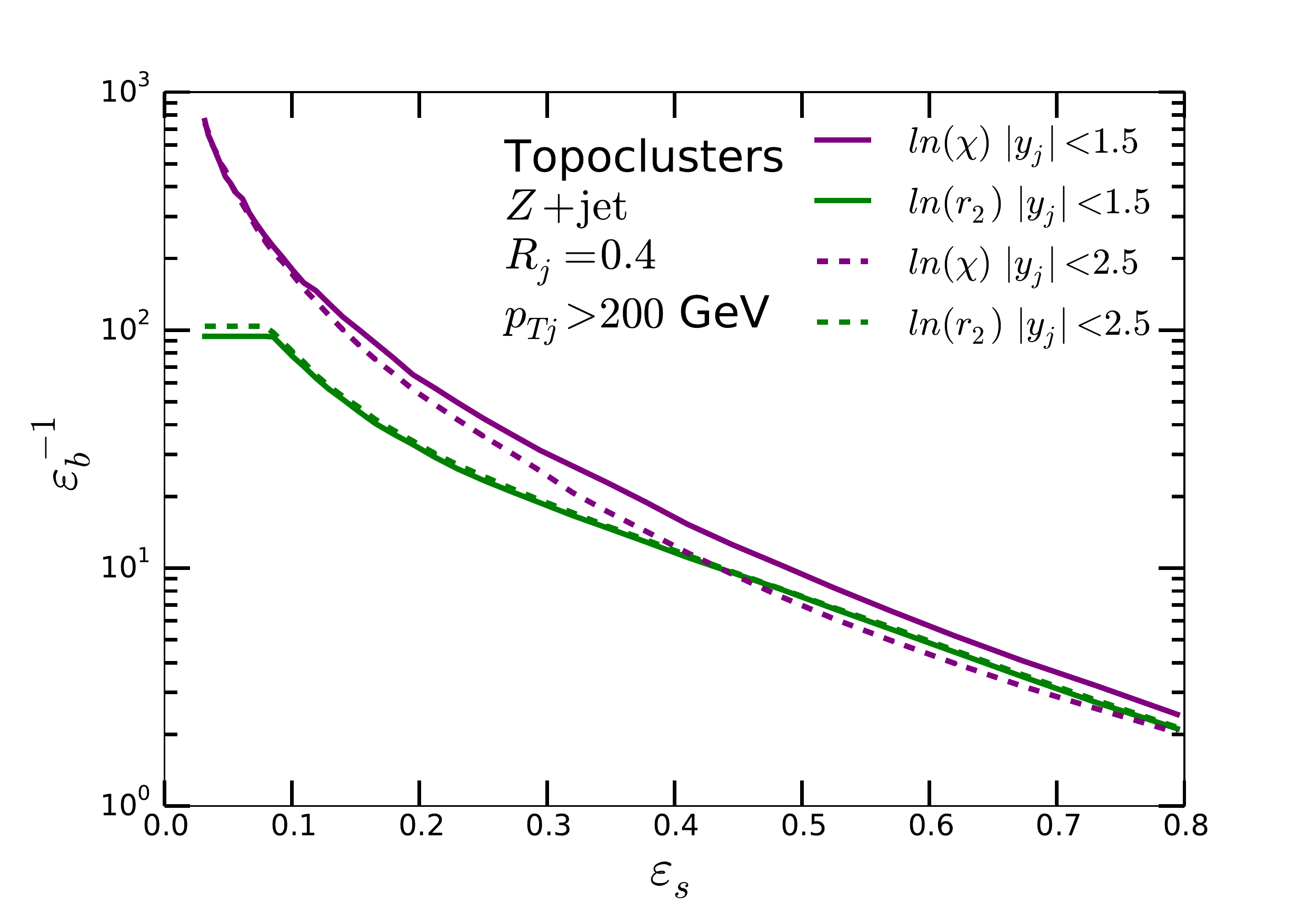}
\includegraphics[width=0.49\linewidth]{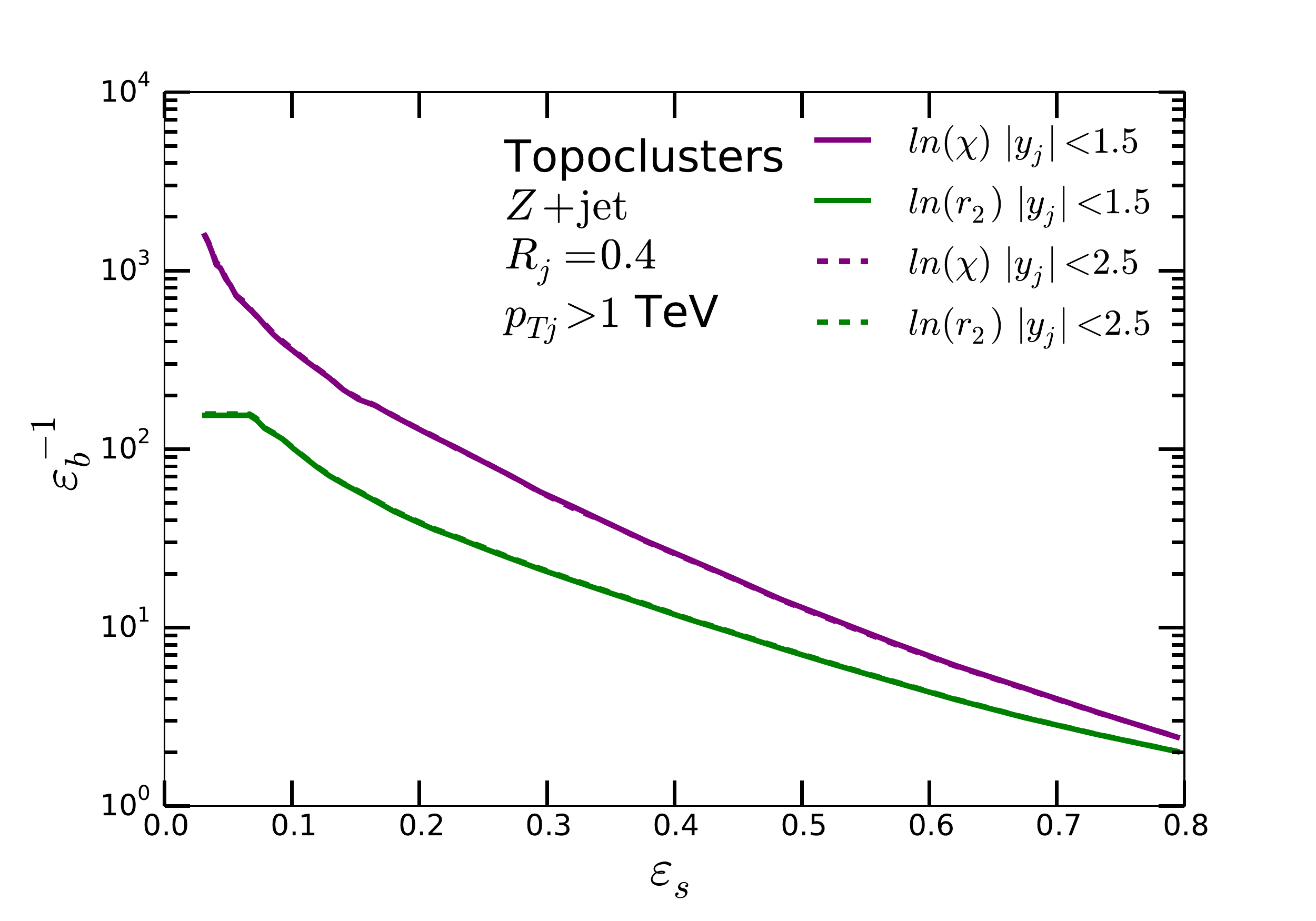}
\caption{Effect of changing the rapidity window. The left panel shows ROC curves for quark tagging and gluon rejection from $Z + \mathrm{jet}$ events for massive topocluster jets with $p_T > 200 \GeV$ for two choices of the rapidity window. The right panel shows the same comparison for $p_T > 1 \TeV$.}
\label{fig:ROC_xZ_ybins_T}
\end{figure}

In the comparisons presented so far, we focused on central jets with rapidity $|y_j|<1.5$. We can ask what happens when we extend the range of jet rapidity to $|y|<2.5$. The results are shown in figure~\ref{fig:ROC_xZ_ybins_T}. For jets with $p_T > 200 \GeV$, the ROC curve for quark tagging using $r_2$ is changed very little when the jet rapidity window is widened. However, ROC curve for quark tagging using $\chi$ becomes worse. This behavior warrants further investigation. If we look at the same question for jets with $p_T > 1 \TeV$, then the effect of widening the rapidity window goes away. This may be because there are not many jets with $p_T > 1\TeV$ and high rapidity.

We next study the effect on quark-gluon discrimination when we increase the radius of the fat jet from $R_{\rm fj} = 0.4$ to $R_{\rm fj} = 0.8$. For the larger fat jet size, we try two versions of shower deconstruction. In the first version, we construct $\chi$ using only one microjet, equal to the fat jet, as we have done in the previous studies with the smaller fat jet size. In the second version, we use the complete shower deconstruction algorithm \cite{Soper:2011cr, Soper:2012pb, Soper:2014rya} as described in section \ref{sec:sd}. The microjets are Cambridge-Aachen jets with $R_{mj}=0.1$ and $p_{Tmj}>5$ GeV. We denote the corresponding likelihood ratio by $\chi^*$. 

\begin{figure}[!h]
\centering 
\includegraphics[width=0.49\linewidth]{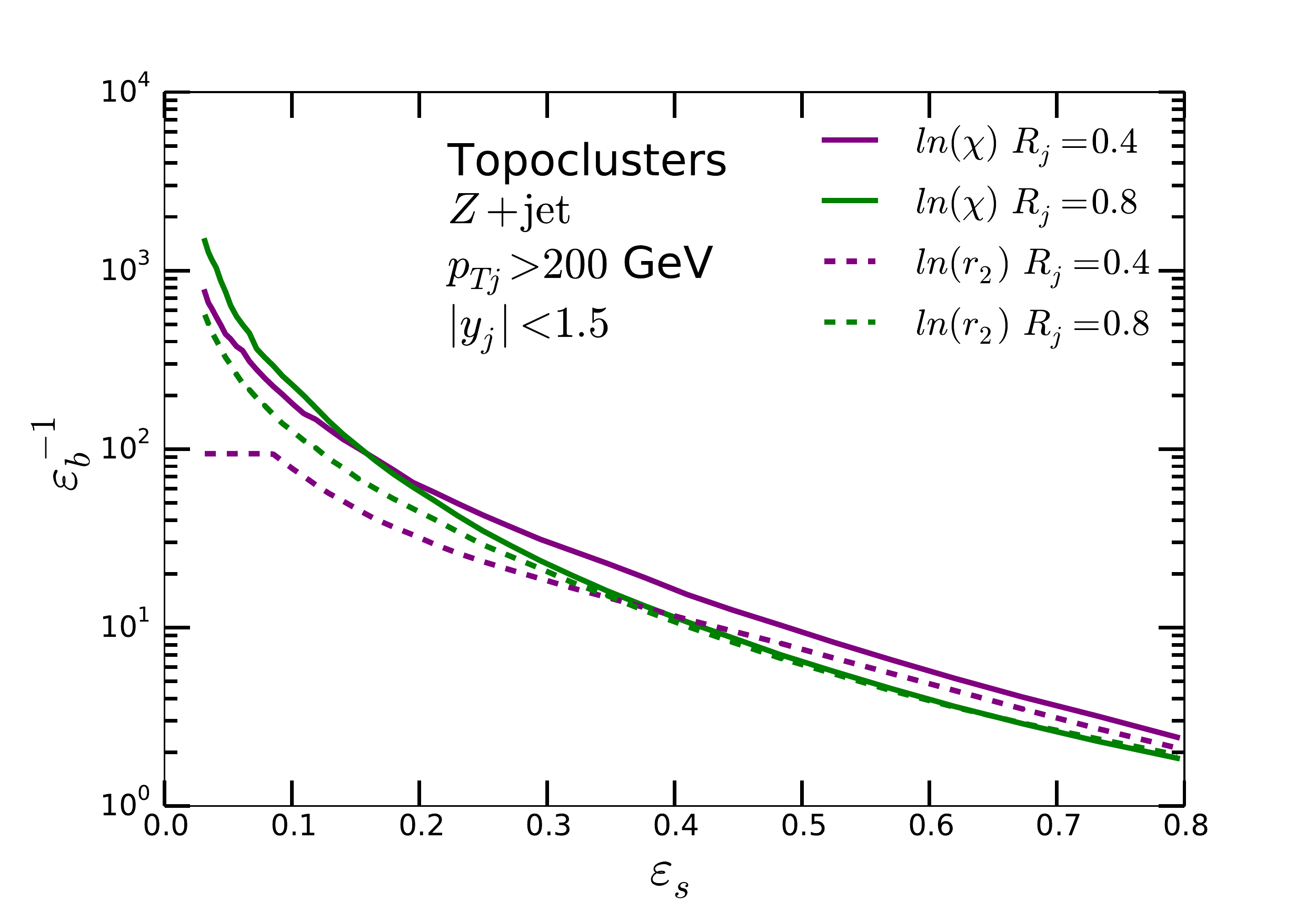}
\includegraphics[width=0.49\linewidth]{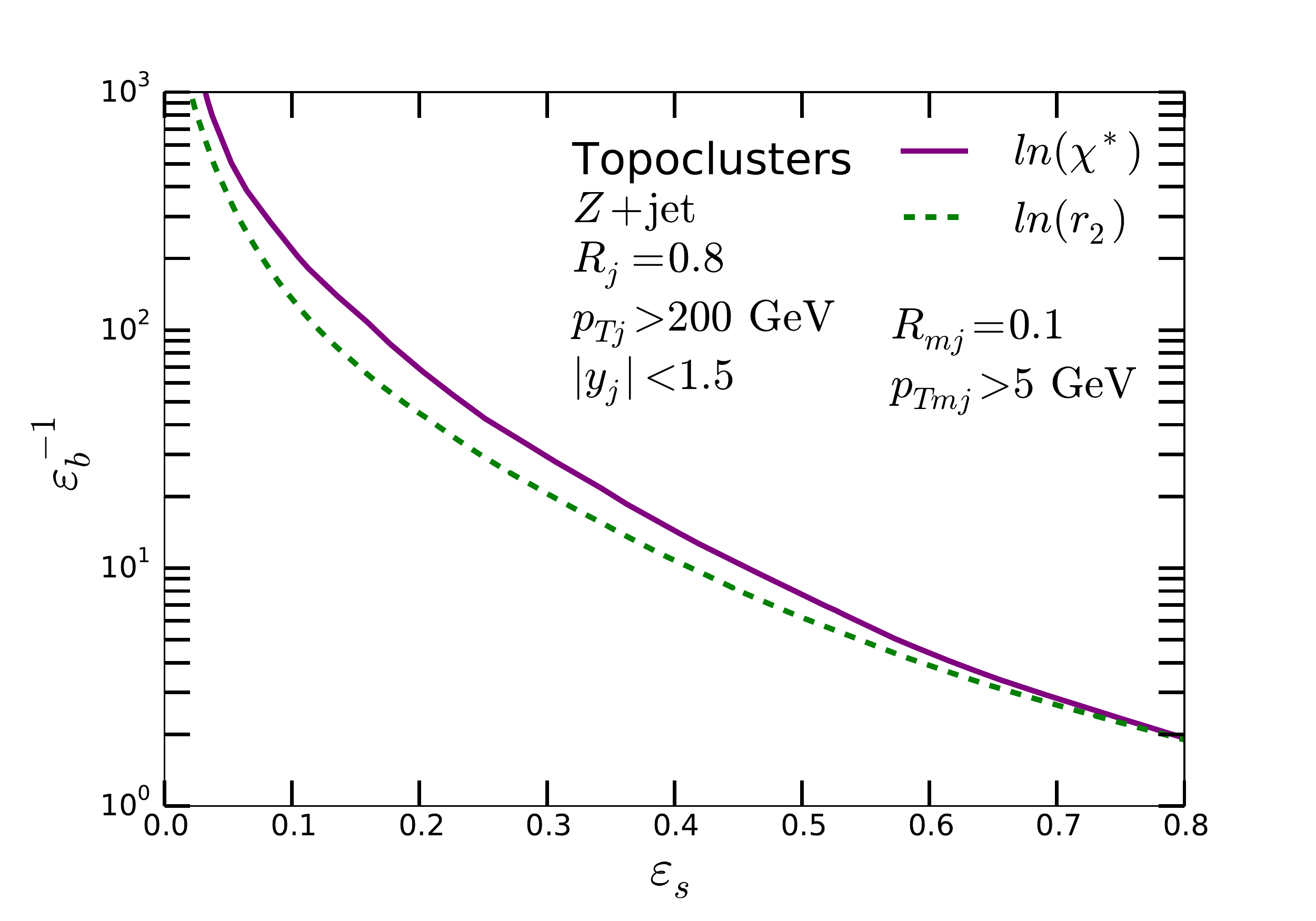}
\caption{ Effect of changing the fat jet radius.
The left panel shows ROC curves for $\ln(\chi)$ and $\ln(r_2)$ from $R=0.4$ and $R=0.8$ Cambridge-Aachen jets built from massive topoclusters. The right panel shows ROC curves from $R=0.8$ jets for $\ln(r_2)$ and full shower deconstruction ($\ln(\chi^{*})$). The microjets for $\chi^*$ are Cambridge-Aachen jets with $R_{mj}=0.1$ and $p_{Tmj}>5$ GeV.}
\label{fig:ROC_SDvR2_MultiR}
\end{figure}

We compare ROC curves for $r_2$ and $\chi$ in in the left plot of figure~\ref{fig:ROC_SDvR2_MultiR}. We see that the ROC curve for $r_2$ improves in the lower half of the $\varepsilon_s$ range and diminishes somewhat in the upper half of the range as the fat jet radius increases. However, for most of the $\varepsilon_s$ range, the ROC curve for the one-microjet version of $\chi$ becomes worse with a fatter fat jet. For $R_{\rm fj} = 0.8$, we compare ROC curves for $r_2$ and $\chi^*$ in right plot of figure~\ref{fig:ROC_SDvR2_MultiR}. We find that full shower deconstruction performs better than $r_2$ across the whole range of signal efficiencies.

\section{Sensitivity to the underlying process and parton shower}
\label{sec:systematics}

If we want to use quark-gluon discrimination in a search for new physics or a measurement of Higgs properties, we need to know the ROC curves for the observables we use as accurately as possible. Otherwise, the measurements will suffer from substantial systematic uncertainty. We can imagine calibrating the ROC curves by comparing experiment to results from event generators for known Standard Model processes. For this to work, we need to be sure that the performance of the observables we use does not depend on the underlying hard process. However, it was shown in \cite{Ask:2011zs,Joshi:2012pu} that jet observables may depend on the event's colour flow. Such a conclusion was reached in \cite{Aad:2014gea, Badger:2016bpw} also for quark and gluon tagging specifically. Thus we need to check whether this is the case for the observables that we have studied.

\begin{figure} [!h]
\centering
\includegraphics[width=0.7\linewidth]{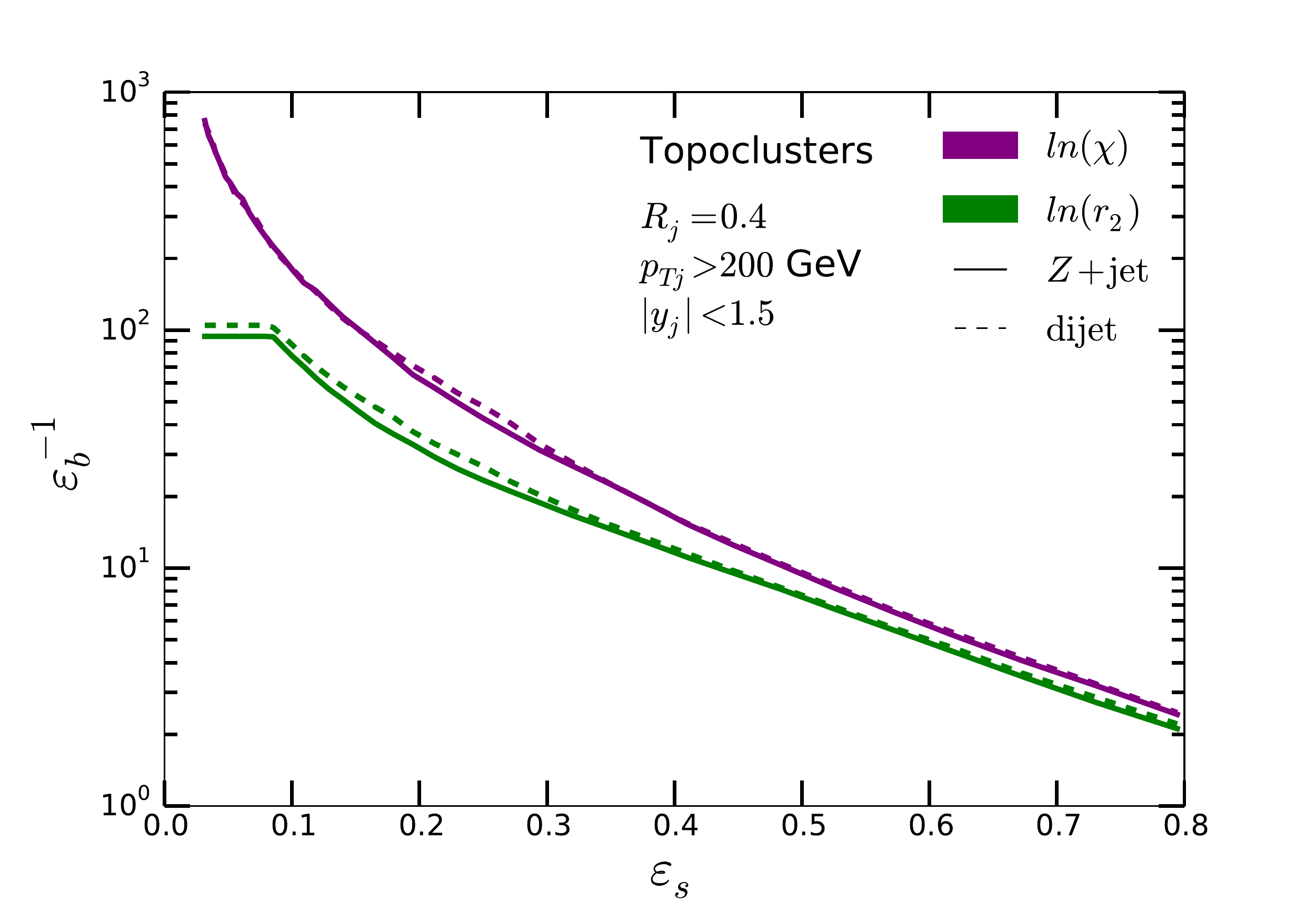}
\caption{ROC curves for $\chi$ and $r_2$ applied to the leading jet of $Z + \mr{jet}$ and $\mr{dijet}$ events.}
\label{fig:ROC_xx_v_xZ}
\end{figure}

In figure~\ref{fig:ROC_xx_v_xZ}, using Pythia 8 events, we compare the $\chi$ ROC curve for tagging quark jets in Z + jet events to that for dijet events. There is hardly any difference. We do the same for $r_2$ and again find hardly any difference. When compared to the difference between the $\chi$ and $r_2$ methods, it becomes evident that quark tagging with either is reliable for jets from different hard processes. Even though we only show the results with a single jet definition, we have confirmed it for jets with larger transverse momentum as well as larger radius parameter.

\begin{figure} [!h]
\centering
\includegraphics[width=0.7\linewidth]{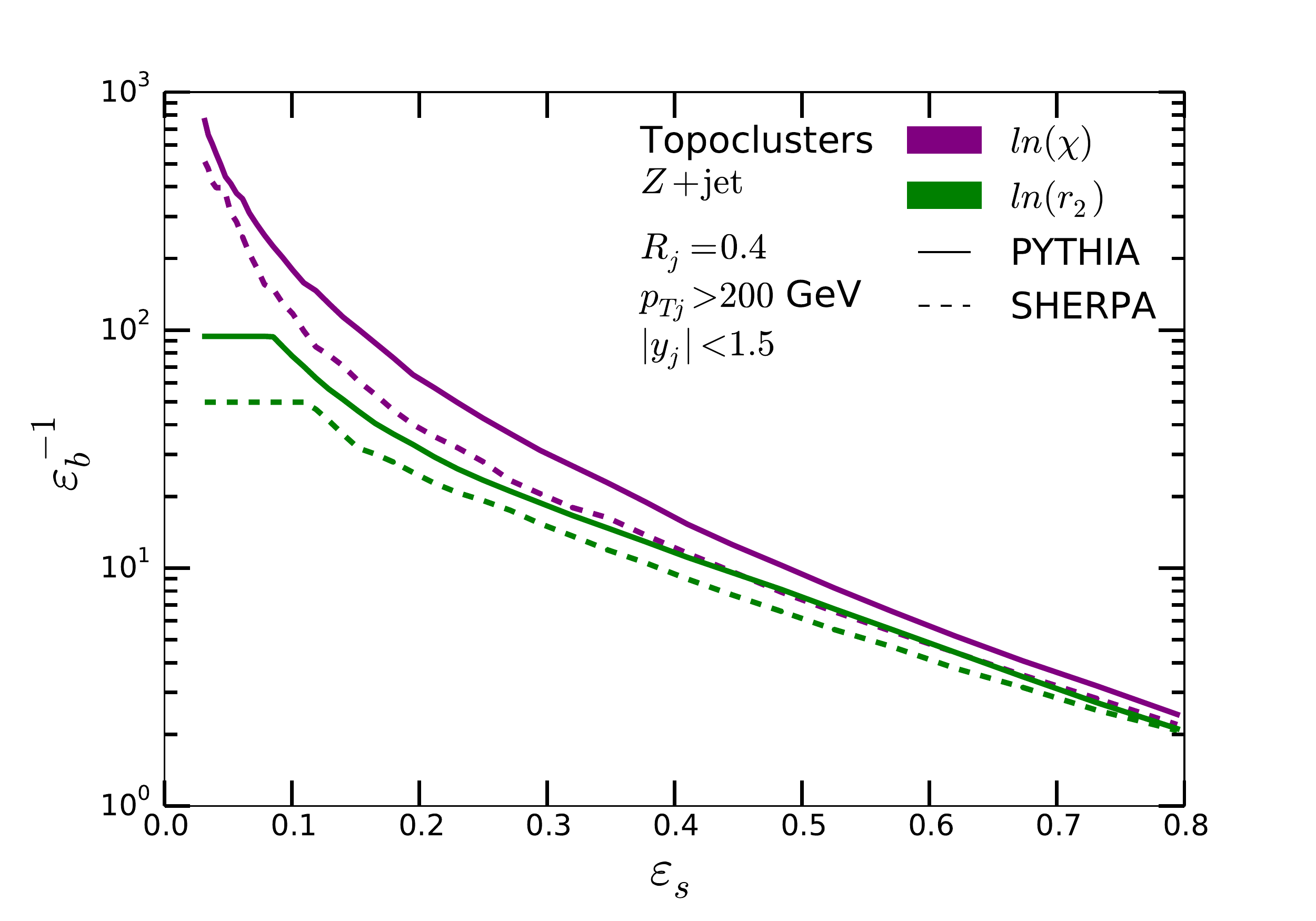}
\caption{ROC curves for $\chi$ and $r_2$ applied to the leading jet of $Z + \mr{jet}$ events generated with Pythia and Sherpa.}
\label{fig:ROC_PvS}
\end{figure}

We can also ask whether existing parton shower Monte Carlos (with their default tunes) are sufficiently accurate to predict the ROC curves for $\chi$ and $r_2$. To answer this question, in figure~\ref{fig:ROC_PvS} we compare the performance of these observables for Z + jet events generated by two different parton showers, Pythia 8 \cite{Sjostrand:2007gs} and Sherpa \cite{Gleisberg:2008ta}.
For $\chi$, we see that there is a rather substantial difference over much of the $\epsilon_s$ range. For $r_2$, the difference is not quite as large, but still not negligible. 

What accounts for this difference? We can look at Section IV.5 of Ref.\cite{Badger:2016bpw} for some insight. The authors of this study looked at quark-gluon discrimination in electron-positron annihilation using generalized angularity observables that are perturbatively infrared safe (and some that are not infrared safe, which we do not discuss here). An infrared safe observable is, by definition, not sensitive to parton splittings that are infinitesimally close to the soft or collinear singularities of perturbation theory. Nevertheless, such an observable can be sensitive to splittings that are at numerically small momentum scales. The study \cite{Badger:2016bpw} examined quark gluon discrimination using several parton shower programs. When hadronization was turned off, there were very substantial differences in quark-gluon discrimination among the programs. It is not clear, at least to us, what characteristics of the parton shower programs led to greater or less quark-gluon discrimination. When hadronization was turned on, quark-gluon discrimination generally increased, suggesting quark jets hadronize quite differently from gluon jets and that this difference affects even nominally infrared safe observables. There were again very substantial differences in quark-gluon discrimination among the programs, but the differences now appeared to depend heavily on the hadronization model that the programs used.

Evidently, if parton shower event generators are to be useful in the analysis of quark-gluon discrimination, they need to better reflect the differences between quark jets and gluon jets, so that the parton shower dependence seen in figure~\ref{fig:ROC_PvS} is reduced. We believe that this goal is achievable. It seems clear that hadronization has an important effect on variables that are sensitive to the difference between quark and gluon jets. The hadronization models in the shower program, as well as certain other parameters in the programs, can be tuned to match data. We note that the mixture of quark and gluon jets inevitably differ between jets in ${\rm p} +  {\rm p} \to {\rm jet} + {\rm jet}$ and ${\rm p} + {\rm p} \to {Z} + {\rm jet}$. Thus, if the data used for Monte Carlo tuning include quark-gluon sensitive observables applied to jets in these two processes, then it seems at least plausible that the tuned shower programs would do better in describing both quark jets and gluon jets.

\section{Application of quark-gluon tagging}
\label{sec:application}

\subsection{Dark matter mono-jet}
\label{sec:monojet}

Searches for dark matter at the LHC have become a vibrant field of research in recent years \cite{Beltran:2010ww,Goodman:2010yf,Fox:2011pm,Abdallah:2014hon}. If the dark matter particle communicates via a mediator with the Standard Model (SM) sector, given a small enough mass of the dark matter candidate, it can be produced at the LHC. While the dark matter particle is only weakly interacting with the detector material, its presence can be inferred indirectly by measuring its associated production with SM particles that carry large transverse momentum, e.g. jets. As shown in \cite{Khachatryan:2014rra}, the dominating backgrounds to high-$p_T$ mono-jet searches are $Z+\mathrm{jet}$ and $W+\mathrm{jet}$. 
Due to the large invariant-mass final state and the structure of parton distribution functions, both of the gauge bosons are likely to be produced in association with a quark rather than a gluon, see table~\ref{tab:monojetXSec}.

Suppose the mediator is a scalar particle that couples to SM particles in agreement with the paradigm of minimal flavor violation, e.g. according to the Lagrangian \cite{Buckley:2014fba,Harris:2014hga}
\begin{equation}
\label{eq:lag}
\mathcal{L}_{\mathrm{scalar}} \supset\, -\,\frac{1}{2}m_{\rm MED}^2 S^2 - g_{\rm DM}  S \, \bar{x}x
 - \sum_q g_{SM}^q S \, \bar{q}q  - m_{\rm DM} \bar{x}x \,.
\end{equation}
The coupling constant $g_{\mathrm{DM}}$ denotes the interaction of the messengers with the dark sector particles. For simplicity we take the dark matter candidate to be a Dirac fermion $x$. The messenger's couplings to quarks are taken to be proportional to the corresponding Higgs Yukawa couplings $y_q=m_q/v$. As a reference and for definiteness we take $g_{\mathrm{DM}}=y_{\mathrm{DM}}$ and $g^q_{\mathrm{SM}}=y_q$. Hence, the mediator couples preferentially to the top quark and decays for large $g_{\mathrm{DM}}$ to dark matter particles. In this case most of the jets produced in association with the dark matter particles are gluon-induced and the signal strength corresponds to the one of the SM Higgs boson with $m_H = 200$ GeV and $\mathrm{BR}(h\to \bar{x}x)\simeq 1$, see table~\ref{tab:monojetXSec}. 

We use Pythia 8 to calculate signal $S+\mathrm{jet}$ and background $Z+\mathrm{jet}$ event rates. We assume the dark matter and mediator masses to be $m_{\mathrm{DM}} = 20$ GeV and $m_{\mathrm{MED}}=200$ GeV respectively.

Even for such an optimistic scenario, the signal-to-background ratio $S/B$ is small, i.e. $S/B \lesssim 0.07$, and systematic uncertainties on measurements , and systematic uncertainties on measurements with missing transverse energy are generically large \cite{Aaboud:2016tnv}. The combined set of uncertainties in this channel, as shown in table~1 of \cite{Khachatryan:2014rra}, amounts to $5 - 10\%$. Hence, a signal-to-background ratio of less than $10\%$ can render this search for cross sections we consider insensitive. Therefore, due to the lack of useful kinematic observables in this simple $2 \to 2$ process, applying a quark/gluon tagger can be vital to improve $S/B$ beyond a necessary, signal cross-section dependent, threshold. After applying cuts on $\chi(g,q)$ corresponding to $50\%$ and $10\%$ we find $S/B\simeq 0.11 $ and $S/B\simeq 0.13$ respectively. To transform this gain in $S/B$ in a sensitivity improvement for dark matter searches, the systematic uncertainties from quark-gluon tagging should be small. This requires to address points raised in section~\ref{sec:systematics} and, more specifically, the design of q/g-tagging approaches that show a stable performance for a wide class of processes.

\begin{table}
\centering
\begin{tabular}{ccccc}
\hline \hline
\multicolumn{5}{c}{$\sigma(\mathrm{jet}+\mathrm{MET})$ {[}fb{]}}                                                                                                                                 \\ \hline
\multicolumn{5}{c}{$13\;\mr{TeV}\;\mr{LHC}$}                                                                                                                                                               \\ \hline 
\multicolumn{1}{c||}{}                                    & \multicolumn{1}{c|}{$p_{T,j} > 250 \mr{GeV}$} & \multicolumn{1}{c|}{$|y| < 1.5$} & \multicolumn{1}{c|}{$\epsilon(\chi(g,q)) \simeq 50\%$} & $\epsilon(\chi(g,q)) \simeq 10\%$ \\ \hline\hline
\multicolumn{1}{c||}{$pp \to (S\to \bar{x}x)j$}           & \multicolumn{1}{c|}{190}                      & \multicolumn{1}{c|}{139}         & \multicolumn{1}{c|}{46.5}               & 8.17               \\ \Xhline{3\arrayrulewidth}
\multicolumn{1}{c||}{$pp \to (S\to \bar{x}x)g$}           & \multicolumn{1}{c|}{96.5}                     & \multicolumn{1}{c|}{78.6}        & \multicolumn{1}{c|}{36.7}               & 6.77               \\ \hline
\multicolumn{1}{c||}{$pp \to (S\to \bar{x}x)q$}           & \multicolumn{1}{c|}{93.3}                     & \multicolumn{1}{c|}{60}          & \multicolumn{1}{c|}{9.27}               & 1.14               \\ \hline \hline
\multicolumn{1}{c||}{$pp \to (Z\to \bar{\nu}\nu)j$}       & \multicolumn{1}{c|}{2830}                     & \multicolumn{1}{c|}{2170}        & \multicolumn{1}{c|}{430}                & 62.2               \\ \Xhline{3\arrayrulewidth}
\multicolumn{1}{c||}{$pp \to (Z\to \bar{\nu}\nu)g$}       & \multicolumn{1}{c|}{334}                      & \multicolumn{1}{c|}{245}         & \multicolumn{1}{c|}{122}                & 24.6               \\ \hline
\multicolumn{1}{c||}{$pp \to (Z\to \bar{\nu}\nu)q$}       & \multicolumn{1}{c|}{2460}                     & \multicolumn{1}{c|}{1890}        & \multicolumn{1}{c|}{299}                & 40.3               \\ \hline \hline
\multicolumn{1}{c||}{$S/B$}                               & \multicolumn{1}{c|}{0.067}                    & \multicolumn{1}{c|}{0.064}       & \multicolumn{1}{c|}{0.11}               & 0.13               \\ \hline
\end{tabular}
\caption{Production cross sections for a top-philic scalar mediator of mass $m_S=200$ GeV that decays predominantly into dark matter, see eq.~\ref{eq:lag}, and the dominant Standard Model background $Z+\mathrm{jet}$ at $\sqrt{s}=13$ TeV.}
\label{tab:monojetXSec}
\end{table}

\subsection{Separation of gluon- and weak boson fusion in $Hjj$}
\label{sec:hjj}

Several ways have been proposed to separate the gluon-fusion from the weak boson-fusion process in dijet associated Higgs production $pp\to Hjj$. Among the methods proposed are rapidity gaps \cite{Dokshitzer:1991he,DelDuca:2001fn}, mini-jet vetos \cite{Rainwater:1999sd,Cox:2010ug}, the matrix element method \cite{Andersen:2012kn} and event shapes \cite{Englert:2012ct}. We add another arrow to the quiver by applying quark-gluon tagging.

\begin{table}[!t]
\begin{center}
 \begin{tabular}{ c || c | c | c }
\hline \hline
\multicolumn{4}{c}{$\sigma(pp\rightarrow Hjj)$ [fb]}	\tabularnewline\hline
			 \multicolumn{4}{c}{13 TeV LHC}		\tabularnewline\hline
		& $p_{T,j}>50~\mathrm{GeV},~\Delta R_{jj} > 2.0$	& $\epsilon(\mathrm{WBF}) \simeq 50\%$	 & $\epsilon(\mathrm{WBF}) \simeq 10\%$  \tabularnewline\hline\hline
WBF $pp \to Hjj$ & 880  & 440 & 91  \tabularnewline\hline\hline
GF $pp \to Hjj$  & 900  & 180 & 15  \tabularnewline\Xhline{3\arrayrulewidth}
GF $pp \to Hqq$  & 22   & 11  & 2.2 \tabularnewline\hline
GF $pp \to Hgg$  & 450  & 61  & 1.8 \tabularnewline\hline
GF $pp \to Hqg$  & 360  & 90  & 8   \tabularnewline\hline\hline
$S/B$            & 0.98 & 2.5 & 6.1 \tabularnewline\hline
\end{tabular}
\caption{LO production cross sections for gluon- and weak boson fusion of a  Higgs boson with mass $m_H=125$ GeV, separated into the respective partonic subprocesses. The two columns on the right show the results after applying a double quark tag with a combined efficiency of $50\%$ and $10\%$ respectively. }
\label{tb:gfcs}
\end{center}
\end{table}

To show the benefit of our approach we calculate the weak boson and the loop-induced gluon-fusion contributions to $pp \to Hjj$. The former allows to measure Higgs-gauge boson couplings and shows very small theoretical uncertainties \cite{Figy:2003nv,Ciccolini:2007ec,Cacciari:2015jma}.

The number of signal events depends on the sum of production processes $p$ and Higgs decay channel $H \to YY$:  
\begin{equation}
	\sigma(H) \times \mathrm{BR}(YY) \sim \left ( \sum_p g_p^2 \right ) \frac{g^2_{HYY}} {\sum_\mathrm{modes} g_i^2},
\end{equation}
assuming no interference between the different production mechanisms, where $g$ denotes the Higgs couplings involved. The sum in the denominator runs over all kinematically accessible decay modes. Hence, the precision in measuring any Higgs boson coupling benefits from separating the production mechanisms.

We generate the events using Sherpa, including the full top loop dependence and require at least two C/A $R=0.4$ jets with $p_{T,j} > 50$ GeV, $|y_j| < 4.5$ and $\Delta R_{jj} \geq 2.0$. After the initial event selection cuts we already find a cross section ratio between gluon and weak boson fusion of $\sim 1$. For this analysis we do not decay the Higgs boson, as this approach can be applied irrespective of the decay mode of interest. Hence, we abstain from considering other Standard Model backgrounds which would depend strongly on the Higgs decay. 

In table~\ref{tb:gfcs} we show by how much this ratio can be improved after applying a double quark tag on the two hardest jets of the event. We find that the gluon fusion contribution can be confidently reduced and even be rendered irrelevant if the WBF rates allow for tight quark tagging.

To give an example how quark-gluon tagging can improve Higgs coupling measurements, we can consider the process $pp \to jj (H\to ZZ^* \to 4l)$. In general this process is not necessarily considered a prime channel to measure the Higgs boson coupling to massive gauge bosons. Although the process is almost free from reducible backgrounds \cite{CMS-PAS-HIG-16-033}, due to efficient cuts on the four and two-lepton systems, the total rate after hard WBF cuts is quite small ($\ll 0.1$ fb). Using quark-gluon tagging allows us to retain a larger cross section while keeping at the same time gluon-fusion induced Higgs production under control. For the branching ratios of the Higgs and Z bosons we assume $\mathrm{Br}(H\to ZZ^*) \simeq 2.62\cdot 10^{-2}$ and $\mathrm{Br}(Z \to l^+l^-) \simeq 0.06$, where $l$ represents electrons and muons. The number of measured events is calculated as
\begin{equation}
N(\mathrm{WBF}) \equiv \epsilon(\mathrm{WBF}) \cdot \sigma(\mathrm{WBF}) \cdot \mathrm{Br}(H\to 4l)  \cdot \mathcal{L},
\end{equation}
and
\begin{equation}
N(\mathrm{GF}) \equiv \epsilon(\mathrm{GF}) \cdot \sigma(\mathrm{GF}) \cdot \mathrm{Br}(H\to 4l)  \cdot \mathcal{L},
\end{equation}
resulting for an integrated luminosity $\mathcal{L}=1000~\mathrm{fb}^{-1}$ in $N(\mathrm{WBF}) \simeq 83$ and $N(\mathrm{GF})\simeq 85$ before applying quark gluon tags on the accompanying jets. After applying quark-gluon tagging, for the working point $\epsilon(\mathrm{WBF}) \simeq 50\%~(10\%)$ of table~\ref{tb:gfcs}, we find $N(\mathrm{WBF}) \simeq 42~(9)$ and $N(\mathrm{GF}) \simeq 17~(1)$. While the application of quark-gluon tags do not improve on $S/\sqrt{S+B}$, for which we find $S/\sqrt{S+B}\simeq 6.4$ before and $S/\sqrt{S+B}\simeq 5.4$ after quark-gluon tagging with $\epsilon(\mathrm{WBF})=50\%$ respectively. However, the combination of measurements including quark gluon tagging at different working points allows to improve the limit setting on deviations from Standard Model Higgs couplings. 

The analytic dependence of the number of observed events on the coupling modifications can be parametrised as
\begin{equation}
 N_{\mathrm{tot}} = \Delta g_{hgg}^2 \Delta g_{hVV}^2  N(\mathrm{GF}) + \Delta g_{hVV}^4 N(\mathrm{WBF}),
\end{equation}
where $\Delta g_i \equiv g_{i,\mathrm{mod}}/g_{i,\mathrm{SM}}$ and we assumed for simplicity that all Higgs-gauge boson couplings are modified the same way, i.e. $\Delta g_{hWW} = \Delta g_{hZZ} = \Delta g_{hVV}$. Note that interference between WBF and GF is highly suppressed \cite{Andersen:2006ag}. 
\begin{figure} [!h]
\centering
\includegraphics[width=0.7\linewidth]{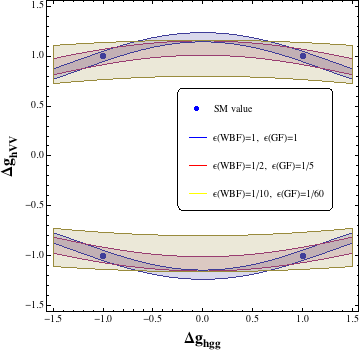}
\caption{Sensitivity bands for the process $pp \to (h \to ZZ^* \to 4l) jj$ after applying quark-gluon tagging with three different working points, assuming a integrated luminosity of $\mathcal{L}=1000~\mathrm{fb}^{-1}$. There is a four-fold ambiguity for the couplings $g_{hVV}$ and $g_{hgg}$, for which the same number of events as in the Standard Model (corresponding to the point $g_{hVV}=1$ and $g_{hgg}=1$) are observed. Coupling modifications are defined as $\Delta g_i \equiv g_{i,\mathrm{mod}}/g_{i,\mathrm{SM}}$.}
\label{fig:hcoup}
\end{figure}

In figure~\ref{fig:hcoup} we show the couplings that can be excluded to roughly 95\% C.L. by requiring $|N_{\mathrm{tot}} - N_{SM}|/\sqrt{N_{SM}} \lesssim 2$. While the sensitivity bands widen for smaller $\epsilon(\mathrm{WBF})$, smaller gluon fusion contributions change the cross section dependence on $g_{hgg}$, thus, increasing sensitivity along otherwise blind directions of coupling combinations. That is, assuming that the experimental results obtained by using three different working points are all consistent with the Standard Model, one can exclude every combination of couplings that is outside of the intersection of the three bands in figure~\ref{fig:hcoup}.

\section{Conclusions}\label{sec:conclusions}

As illustrated in section~\ref{sec:application}, tagging jets as being likely quark initiated or likely gluon initiated can be used for separating signal from background in LHC events. In the earlier sections of this paper, we studied issues related to how such quark-gluon tagging can be performed.

Our studies suggest that, at least for the methods investigated, quark-gluon tagging can be effective, but has a substantial sensitivity to physics at rather small momentum scales. This is illustrated by the finding in figure~\ref{fig:ROC_PvS} that if we seek to tag quark jets, then the background rejection factors obtained with events generated by standard Monte Carlo event generators differ according to which generator, Pythia or Sherpa, we use. The ROC curves obtained are qualitatively similar but have significant quantitative differences. Another finding, illustrated in figures~\ref{fig:ROC_xZ_y15_R2vC1_Rfixed} and \ref{fig:ROC_xZ_y15_R2vC1_pTfixed}, that points to the same conclusion is that different results are obtained by examining the jet substructure beginning with hadrons or beginning with simulated massive topoclusters. Starting with hadrons gives the most detailed view, while starting with topoclusters removes some of the information that comes from the final, infrared dominated, stages of hadronization. What we see is that including or not this infrared dominated information affects the results.

This tentative conclusion suggests that there is a tradeoff in using quark-gluon tagging between sensitivity to the signals that we are looking for and the reliability of the method. That is, we can improve background rejection and thus increase our chances of finding, say, a signal for new physics. However, we may induce a substantial systemic error in the calculation of the amount of background rejection. Of course, if we can measure the background rejection factor experimentally, this problem is ameliorated. To this end, it is encouraging that, when we try to tag quark jets, the background rejection factor seems to be quite independent of the hard scattering process that creates the jets, as illustrated in figure~\ref{fig:ROC_xx_v_xZ}.

We examined several measures of jet substructure that bear on quark-gluon separation. The most realistic case is to apply these measures to simulated topoclusters rather than hadrons, both because topocluster results are likely to be less infrared sensitive and because they are more experimentally practical. In our studies, we retained the mass of each simulated topocluster rather than scaling the momentum so as to set the topocluster mass to zero. This goes beyond the method used by ATLAS, but it improves the quark-gluon separation for the shower deconstruction variable $\chi$. Most of our studies concerned tagging fat jets with radius parameter $R_{\rm fj} = 0.4$. There we found that the variables $r_1$, $r_2$ and $C_1$ exhibited similar performances, as illustrated in figure~\ref{fig:ROC_AllVars}. For other graphs, we chose $r_2$ as representative of these three. We compared $r_2$ to the shower deconstruction variable $\chi$. Normally, shower deconstruction divides the fat jet into several smaller jets, called microjets. That is essential when seeking to find heavy particles that decay to several jets. However, in distinguishing quark from gluon QCD jets with a rather small cone size $R_{\rm fj} = 0.4$ for the fat jet, we found that it was better to simply apply the shower deconstruction calculation of $\chi$ to a single microjet, identical to the fat jet. The result, from figure~\ref{fig:ROC_AllVars}, is that the ROC curve for $\chi$ shows better background rejection than that for $r_2$.

We examined quark-gluon discrimination also for fatter fat jets, with $R_{\rm fj} = 0.8$, as illustrated in figure \ref{fig:ROC_SDvR2_MultiR}. There we found that the shower deconstruction method with more than one microjets worked best. However, the improvement over the use of $R_{\rm fj} = 0.4$ fat jets was small.

We conclude, in general agreement with refs. \cite{Catani:1992jc,Thaler:2010tr,Gallicchio:2011xq,Larkoski:2013eya,Larkoski:2014pca,Bhattacherjee:2015psa,Badger:2016bpw}, that using jet substructure measures to discriminate between quark initiated jets and gluon initiated jets can be helpful for distinguishing signals from backgrounds at the LHC. We have presented results that bear on the use of these methods, but a final judgement can only be reached by using these observables by ATLAS and CMS.

\vskip 1 \baselineskip

\noindent {\it{Acknowledgments. We thank Silvan Kuttimalai for support with Sherpa event generation. MS is supported in part by the European Commission through the ``HiggsTools'' Initial Training Network PITN-GA-2012-316704. DS is supported by the U.S.\ Department of Energy. DFL is supported by the Alexander von Humboldt Foundation.}}

  
\bibliographystyle{JHEP}
\bibliography{references}

\end{document}